\documentclass[aps,prl,superscriptaddress,twocolumn,10pt]{revtex4-2}

\usepackage[centering,hmargin=1.7cm,vmargin=0.78in]{geometry}

\usepackage{amsmath,mathtools,amsthm,amssymb}
\usepackage{tabularx}
\usepackage{tabularray}

\usepackage{graphicx}

\usepackage{color}
\usepackage{bbold}
\usepackage{enumitem}
\usepackage{physics}
\usepackage{comment}
\usepackage{array}
\usepackage{graphics}
\usepackage{wrapfig}
\usepackage{fontsize}
\graphicspath{{figures/}}
\usepackage{bbm}
\usepackage{dsfont}
\usepackage{mathrsfs}
\usepackage{mathdots}
\usepackage{enumitem}
\usepackage{pifont}
\usepackage[table]{xcolor}
\usepackage{booktabs}
\usepackage{pifont}
\usepackage{amssymb}
\newcommand{\cmark}{\ding{51}}%
\newcommand{\xmark}{\ding{55}}%
\definecolor{lightred}{RGB}{243,229,231}
\definecolor{lightgreen}{RGB}{241,255,239}
\definecolor{lightblue}{RGB}{232,240,244}
\definecolor{RoyalBlue}{RGB}{65,105,225}
\definecolor{ForestGreen}{RGB}{34,139,34}   
\definecolor{Maroon}{RGB}{135,0,0}
\definecolor{myrefcolor}{rgb}{0.067,0.5,0.5}
\definecolor{myurlcolor}{rgb}{0.1,0,0.9}

\usepackage[
    breaklinks,
    pdftex,
    colorlinks=true,
    linkcolor=myrefcolor,
    citecolor=myrefcolor,
    urlcolor=myrefcolor
]{hyperref}
\usepackage[capitalize]{cleveref}

\DeclareDocumentCommand\mel{ s s m m m }
{ 
    \IfBooleanTF{#1}
    {
        \IfBooleanTF{#2}
        {\left\langle{#3}\middle\vert{#4}\middle\vert{#5}\right\rangle} 
        {\vphantom{#3#4#5}\left\langle\smash{#3}\middle\vert\smash{#4}\middle\vert\smash{#5}\right\rangle} 
    }
    {\vphantom{#4}\left\langle{#3}\middle\vert\smash{#4}\middle\vert{#5}\right\rangle} 
}

\newtheorem*{theorem*}{Theorem}
\newtheorem{theorem}{Theorem}
\newtheorem{lemma}{Lemma}

\newtheorem{proposition}{Proposition}
\newtheorem{definition}{Definition}

\newtheorem{corollary}{Corollary}

\theoremstyle{remark}

\usepackage{algorithm}
\usepackage{algpseudocode}

\newcommand{\be}{\begin{equation}\begin{aligned}\hspace{0pt}}
\newcommand{\ee}{\end{aligned}\end{equation}}
\newcommand{\ba}{\begin{eqnarray}}
\newcommand{\ea}{\end{eqnarray}}

\definecolor{airforceblue}{rgb}{0.36, 0.54, 0.66}

\newcommand{\bb}{\begin{equation}\begin{aligned}\hspace{0pt}}
\newcommand{\bbb}{\begin{equation*}\begin{aligned}}
\newcommand{\eb}{\end{aligned}\end{equation}}
\newcommand{\eeb}{\end{aligned}\end{equation*}}
\allowdisplaybreaks
\begin{document}

\title{Stabilizer entropies are monotones for magic-state resource theory}



\author{Lorenzo Leone}
\thanks{Contributed equally. \{\href{mailto:lorenzo.leone@fu-berlin.de}{lorenzo.leone}, \href{mailto:l.bittel@fu-berlin.de}{l.bittel}\}@fu-berlin.de}
\affiliation{Dahlem Center for Complex Quantum Systems, Freie Universit\"at Berlin, 14195 Berlin, Germany}
\author{Lennart Bittel}
\thanks{These authors contributed equally. \{\href{mailto:lorenzo.leone@fu-berlin.de}{lorenzo.leone}, \href{mailto:l.bittel@fu-berlin.de}{l.bittel}\}@fu-berlin.de}
\affiliation{Dahlem Center for Complex Quantum Systems, Freie Universit\"at Berlin, 14195 Berlin, Germany}


\begin{abstract}
Magic-state resource theory is a powerful tool with applications in quantum error correction, many-body physics, and classical simulation of quantum dynamics. Despite its broad scope, finding tractable resource monotones has been challenging. Stabilizer entropies have recently emerged as promising candidates (being easily computable and experimentally measurable detectors of nonstabilizerness) though their status as true resource monotones has been an open question ever since. In this Letter, we establish the monotonicity of stabilizer entropies for $\alpha \geq 2$ within the context of magic-state resource theory restricted to pure states. Additionally, we show that linear stabilizer entropies serve as strong monotones. Furthermore, we extend stabilizer entropies to mixed states as monotones via convex roof constructions, whose computational evaluation significantly outperforms optimization over stabilizer decompositions for low-rank density matrices. As a direct corollary, we provide improved conversion bounds between resource states, revealing a preferred direction of conversion between magic states. These results conclusively validate the use of stabilizer entropies within magic-state resource theory and establish them as the only known family of monotones that are experimentally measurable and computationally tractable.
\end{abstract}

\maketitle

{\em Introduction.---} The inception of magic-state resource theory can be traced back to the seminal work of Bravyi and Kitaev~\cite{bravyi_universal_2005}. They showed the feasibility of applying non-Clifford gates to a target state through \textit{stabilizer operations} -- i.e., Clifford operations and measurements -- leveraging a single-qubit auxiliary state prepared in a nonstabilizer (or ``magic'') state. Building upon this fundamental result, magic-state resource theory was then formalized~\cite{Veitch_2014}. The core principle hinges on the following dichotomy: while stabilizer operations are easily implemented in a fault-tolerant fashion (i.e., transversally) in many schemes~\cite{campbell_bound_2010}, achieving fault-tolerance for nonstabilizer operations proves to be challenging~\cite{PhysRevLett.102.110502}. As a result, stabilizer operations are regarded as the \textit{free operations} within the resource theory framework, with stabilizer states designated as the corresponding \textit{free states}, while non-Clifford operations and nonstabilizer states are considered \textit{resourceful}. The task at hand involves utilizing nonstabilizer resource states and exclusively employing stabilizer operations to implement non-Clifford unitary gates fault-tolerantly, a process commonly referred to as \textit{magic-state distillation}.

Stabilizer operations are not only non-universal, but have also been shown to be efficiently classically simulatable~\cite{gottesman_heisenberg_1998}. Nonstabilizer states become hard to simulate classically~\cite{aaronson_improved_2004}. Hence, beyond the scope of magic-state distillation, magic-state resource theory is also employed to quantify the hardness of classically simulating quantum states using stabilizer formalism. From this perspective, nonstabilizer states are regarded as resourceful since they enable quantum computations that surpass the capabilities of classical computers.

From the technical point of view, the construction of a quantum resource theory is straightforward; given the set of free states $\mathcal{T}$ uniquely defining the resource theory, free operations are quantum channels (completely positive, trace preserving and linear operations on a quantum system) leaving $\mathcal{T}$ invariant. The challenge lies in selecting a \textit{resource monotone} {(or measure)}, akin to a ``thermometer'', to accurately quantify the resource amount in a given state $\psi\not\in\mathcal{T}$, represented by  a positive scalar function. While universal resource monotones (independent from the specific resource theory) do exist~\cite{chitambar_quantum_2019}, they are impractical to compute and experimentally measure in real scenarios. For a resource theory to be applicable in practice, such as in quantum many-body physics and quantum computation, finding a resource-specific measure is an extremely important and nontrivial task.

Stabilizer entropies (SEs) indexed by a $\alpha$-R\'enyi index were recently proposed to probe nonstabilizerness in multiqubit {pure} quantum states~\cite{leone_stabilizer_2022}. They have garnered particular attention due to their analytical~\cite{passarelli2024nonstabilizerness,PhysRevA.106.042426,haug_quantifying_2023,chen2023magic2} and numerical~\cite{lami_quantum_2023,tarabunga2024nonstabilizerness} computability as well as experimental measurability~\cite{oliviero2022MeasuringMagicQuantum,haug_efficient_2023}. Thanks to their computability advantages, stabilizer entropies have significantly advanced the study of magic-state resource theory within the context of many-body physics, allowing numerical simulations up to $\sim 100$ qubits~\cite{lami_quantum_2023,tarabunga2024nonstabilizerness}, which was previously limited to only a few bodies~\cite{heinrich_robustness_2019}.
In addition, through stabilizer entropies, magic-state resource theory found connection to a number of topics in quantum physics, including quantum chaos and scrambling~\cite{leone_stabilizer_2022}, as well as in quantum information with applications in cryptography~\cite{gu2023little}, fidelity estimation~\cite{leone_nonstabilizerness_2023}, Pauli sampling~\cite{hinsche2024efficient,leone2023learning}, and classical shadows~\cite{zhang2024minimal,brieger2023stability}.

However, a counterexample to the monotonicity of stabilizer entropies with R\'enyi index $\alpha<2$ under stabilizer protocols was provided~\cite{haug_stabilizer_2023}. Since then, the question of whether stabilizer entropies with $\alpha\geq 2$, when restricted to pure states, serve as a monotone for magic-state resource theory has remained unanswered. Given the significance of stabilizer entropies across various contexts, the question ``are stabilizer entropies good monotones for magic-state resource theory?'' has garnered increasing attention.

In this work, we demonstrate that stabilizer entropies with integer R\'enyi index $\alpha\geq 2$ are monotones for magic-state resource theory when restricted to pure states. We also show that their linearized versions -- linear stabilizer entropies -- serve as \textit{strong} monotones. {Moreover, we extend stabilizer entropies to mixed states through convex-roof construction, ensuring monotonicity under stabilizer protocols and offering substantial computational advantages over known monotones for low-rank density matrices.} See \cref{table1} for a summary. {As an application of our results, we improve over state-of-the-art resource conversion bounds.} These results conclusively validate the use of stabilizer entropies to quantify nonstabilizerness in multiqubit quantum systems.
\begin{table}[!h]
  \centering
  \begin{tabular}{@{}|l|c|c|@{}}
    \hline
    & \textbf{Stabilizer monotone} & \textbf{Strong stabilizer monotone} \\
    \hline
    $M_{\alpha<2}$&    \cellcolor{lightred} \textcolor{Maroon}{\xmark}    & \cellcolor{lightred}\textcolor{Maroon}{\xmark} \\
   $M_{\alpha\ge 2}$&
   \cellcolor{lightgreen}\textcolor{ForestGreen}{\cmark} & \cellcolor{lightred}\textcolor{Maroon}{\xmark} \\
  $M_{\alpha<2}^{\mathrm{lin}}$ &  \cellcolor{lightred}\textcolor{Maroon}{\xmark} &
    \cellcolor{lightred}\textcolor{Maroon}{\xmark} \\ $M_{\alpha\ge 2}^{\mathrm{lin}}$ & \cellcolor{lightgreen}\textcolor{ForestGreen}{\cmark} & \cellcolor{lightblue}\textcolor{RoyalBlue}{\cmark} \\
    \hline
  \end{tabular}
    \caption{Summary of the results. $M_{\alpha}$ and $M_{\alpha}^{\mathrm{lin}}$ represent the stabilizer entropy and the linear stabilizer entropy respectively (see \cref{def:stabilizerentropy}). Both are extended to mixed states. }\label{table1}
\end{table}

{\em Setup.---} Throughout the Letter we consider the Hilbert space $\mathcal{H}_n$ of $n$ qubits and denote as $d_n=2^n$ its dimension. A natural operator basis is given by Pauli operators $P\in\mathbb{P}_n$, i.e. $n$-fold tensor products of ordinary Pauli matrices $I,X,Y,Z$. The subgroup of unitary matrices that maps Pauli operators to Pauli operators is known as the Clifford group. Stabilizer states, denoted as $\ket{\sigma}$ in this work, are pure states obtained from $\ket{0}^{\otimes n}$ with the action of unitary Clifford operators. Throughout the work, we will use $\psi,\phi$ to denote pure states, while $\rho$ to denote (possibly mixed) general states.

{\em Magic-state resource theory.---}   A resource theory is uniquely characterized by its set of free states. The set of free states for magic-state resource theory is given by the convex hull of pure stabilizer states, hereby denoted as $\mathrm{STAB}\coloneqq\{\sum_ip_i\ketbra{\sigma_i}\,:\, p_i\ge 0,\sum_i p_i=1\}$. 

Free operations are the \textit{stabilizer protocols}, that are completely positive and trace preserving maps that leave invariant the set $\mathrm{STAB}$. Denoting as $\mathcal{S}$ the set of stabilizer protocols, every operation $\mathcal{E}\in\mathcal{S}$ can be built out of the following elementary operations: (i) Clifford unitaries; (ii) partial trace; (iii) measurements in the computational basis; (iv) composition with auxiliary qubits in $\ket{0}$;  the above operations conditioned on (v) measurement outcomes and (vi) classical randomness. Since these operations (and combination thereof) return different states based on conditioned operations and non-deterministic strategies, the most general form of a stabilizer protocol applied on a state $\rho$ is $ \mathcal{E}(\rho)=\{(p_i,\rho_i)\}$, i.e. a collection of $n_i$ qubit states $\rho_i$ with probabilities $p_i$. A stabilizer protocol $\mathcal{E}$ is \emph{deterministic} if $\mathcal{E}(\rho)=\tilde{\rho}$, i.e. a unique quantum state is reached with unit probability.

Given stabilizer protocols, one can define monotones for magic-state resource theory.
\begin{definition}[Stabilizer monotone]\label{def:magicmonotone} A stabilizer monotone $\mathcal{M}$ is a real-valued function for all $n\in \mathbb{N}$ qubit systems (or collection thereof) such that: (i) $\mathcal{\mathcal{M}}(\rho)=0$ if and only $\rho\in\mathrm{STAB}$; (ii) $\mathcal{M}$ is nonincreasing under stabilizer protocols $\mathcal{S}$, i.e.
\be
\mathcal{M}(\mathcal{E}(\rho))\le \mathcal{M}(\rho),\quad \forall \mathcal{E}\in\mathcal{S}\,.
\ee

\end{definition}
{As is customary in resource theories~\cite{chitambar_quantum_2019}, there are functions that always serve as monotones regardless of the specific resource theory, which includes (but is not limited to) robustness measures (robustness of magic~\cite{howard_application_2017}), resource-rank measures (stabilizer rank~\cite{bravyi_improved_2016}) or distance-based measures (relative entropy of magic~\cite{liu_manybody_2022}).} While all of them require an impractical optimization over the set of free states (stabilizer decomposition), the challenge has always been to find a tractable resource monotone that meaningfully apply to the specific resource theory at hand, e.g., von Neumann entanglement entropy for entanglement. In magic-state resource theory, such a role is played by the stabilizer R\'enyi entropy~\cite{leone_stabilizer_2022}. As we shall see, the stabilizer entropy is naturally defined for pure states. We thus define a monotone for magic-state resource theory restricted to pure states.
\begin{definition}[Pure-state stabilizer monotone]\label{def:purestatemonotone} A pure-state stabilizer monotone is a real-valued function for all $n\in\mathbb{N}$ qubit systems such that: (i) $\mathcal{M}(\psi)=0$ if and only if $\ket{\psi}$ is a pure stabilizer state; (ii) for every pair $(\ket{\psi},\mathcal{E})$, where $\mathcal{E}\in\mathcal{S}$ a stabilizer protocol, obeying $\mathcal{E}(\ketbra{\psi})=\ketbra{\phi}$, it holds that:
\be
\mathcal{M}(\phi)\le\mathcal{M}(\psi)\,.
\ee
\end{definition}
In other words, a pure-state stabilizer monotone is a monotone when restricted to \textit{deterministic} pure-state  stabilizer protocols, i.e., those mapping pure states to pure states. Note that, as stabilizer protocols consist of multiple elementary operations, the aforementioned definition does not preclude intermediate states from becoming mixed, provided the final state remains pure. Clearly, being a pure-state monotone is a weaker condition than being a monotone and, as a matter of fact, \cref{def:magicmonotone} implies \cref{def:purestatemonotone}. {However, deterministic pure-state stabilizer protocols have their own relevance: they are mainly employed for the conversion between resource states, which is crucial in the field of quantum error correction for implementing nonstabilizer operations fault-tolerantly~\cite{beverland_lower_2020}.} A famous example is provided by the \textit{magic-state injection}, i.e., $\mathcal{E}(\ket{T}\otimes \ket{\psi})=\ket{0}\otimes T\ket{\psi}$ where $\ket{T}=T\ket{+}$ is the $T$-state and $T\coloneqq \mathrm{diag}(1,e^{i\pi/4})$ being the $T$-gate~\cite{Zhou_2000}. 

As a last notion, let us introduce the concept of strong monotonicity for pure states. An analogous definition for mixed states follows.
\begin{definition}[Strong pure-state stabilizer monotone]\label{def:strongmagicmonotone} Let $\mathcal{M}$ a pure-state stabilizer monotone. Let $\ket{\psi}$ be a pure state and consider the collection of pure states $\mathcal{E}(\psi)=\{(p_i,\ket{\phi_i})\}$, obtained after applying a stabilizer protocol $\mathcal{E}\in\mathcal{S}$. $\mathcal{M}$ is a strong pure-state stabilizer monotone if
\be
\mathcal{M}(\psi)\ge \sum_i p_i \mathcal{M}(\phi_i)\,.
\ee
\end{definition}
Informally, the notion of strong monotonicity says that, on average, the nonstabilizer resource cannot increase after non-deterministic stabilizer protocols. {Operationally, strong stabilizer monotones bound the optimal probability of conversion between resource states via stabilizer operations~\cite{chitambar_quantum_2019}, as explained in greater detail below.} Now that we have all the necessary definitions and tools, the next section reveals the main findings of our work.

{\em Monotonicity of stabilizer entropies.--- } In this section, we establish the monotonicity of stabilizer entropies under stabilizer protocols in various flavors. Let us first define SEs. Given a pure state $\ket{\psi}$, the quantity $d^{-1}_{n}|\langle\psi|P|\psi\rangle|^{2}$ forms a probability distribution on the Pauli group $\mathbb{P}_n$, known also as \textit{characteristic distribution}~\cite{zhu_clifford_2016}. SEs are defined, up to an offset, as $\alpha$-R\'enyi entropies of this probability distribution. 
\begin{definition}[$\alpha$-stabilizer entropy]\label{def:stabilizerentropy} The $\alpha$-R\'enyi stabilizer entropy of a pure quantum state $\ket{\psi}$ on $n$ qubits reads
\be
M_{\alpha}(\psi)\coloneqq\frac{1}{1-\alpha}\log_2 P_{\alpha}(\psi),\, P_{\alpha}(\psi)\coloneqq\frac{1}{d_n}\sum_{P\in\mathbb{P}_n}|\langle\psi|P|\psi\rangle|^{2\alpha}\!,\nonumber
\ee
where $P_{\alpha}(\psi)$ is referred to as stabilizer purity. Linear stabilizer entropies are defined as $M_{\alpha}^{\mathrm{lin}}(\psi)\coloneqq 1-P_{\alpha}(\psi)$.
\end{definition}
Stabilizer entropies exhibit two key properties required for magic-state resource theory: (i) they are minimized, i.e., are zero, only for stabilizer states and (ii) are invariant under the action of Clifford unitary operators. {Additionally, SEs possess several useful properties, such as additivity under tensor products, which will be crucial for bounding resource conversion rates, as explained later.} Moreover, they serve as lower bounds for key monotones within magic-state resource theory, including the robustness of magic~\cite{howard_application_2017},  min-relative entropy of magic~\cite{liu_manybody_2022}, stabilizer nullity~\cite{beverland_lower_2020} and stabilizer extent~\cite{Bravyi_2019}. For a comprehensive list, please refer to Ref.~\cite{_see_}. 

However, mere invariance under Clifford operators does not suffice to establish stabilizer entropies as effective monotones for magic-state resource theory, given the significantly broader scope of stabilizer protocols. {In fact, in a recent study~\cite{haug_stabilizer_2023}, a specific counterexample to deterministic pure-state stabilizer protocols involving Clifford operations conditioned on measurement outcomes has been presented for $\alpha<2$.} Whether stabilizer entropies serve as stabilizer monotones for larger values of $\alpha$ has remained an open question, and given the extensive list of useful features, it is indeed significant. We demonstrate below that stabilizer entropies function as monotones for deterministic stabilizer protocols for any $\alpha\ge2$, thereby qualifying as pure-state stabilizer monotones according to \cref{def:purestatemonotone}.
 
\begin{theorem} Stabilizer entropies $M_{\alpha}$ are pure-state stabilizer monotones for every integer $\alpha\ge2$.\label{th:stabentropypuremonotones}
\end{theorem}
\noindent
{\em Proof sketch.} First of all, it is crucial to note that a deterministic pure-state stabilizer protocol only necessitates the initial and final states to be pure. We consider the following decomposition of a pure state $\ket{\psi}=\sqrt{p}\ket{0}\otimes\ket{\phi_1}+\sqrt{1-p}\ket{1}\otimes\ket{\phi_2}$. The key step of the proof is showing that $M_{\alpha}(\psi)\ge \min\{M_{\alpha}(\phi_1),M_{\alpha}(\phi_2)\}$. From this simple observation, for a general stabilizer protocol $\mathcal{E}\in\mathcal{S}$, it holds that
\be M_{\alpha}(\psi)\ge \inf_{(q_{ij},\phi_{ij})}\Big\{\min_{ij}M_{\alpha}(\phi_{ij})\,:\, p_i\rho_i =\sum_jq_{ij} \phi_{ij}\Big\}\,,
\nonumber\ee
where the infimum is over the possible pure state convex decompositions of $p_i\rho_i$ where $\mathcal{E}(\psi)=\{(p_i,\rho_i)\}$.  By requiring $\mathcal{E}$ to be a deterministic pure-state stabilizer protocol, the result just follows. The complete proof can be found in~\cite{_see_}.\qed

In fact, we can present an even stronger result. We show that linear stabilizer entropies are strong monotones according to \cref{def:strongmagicmonotone}.

\begin{theorem} Linear stabilizer entropies $M_{\alpha}^{\mathrm{lin}}$ are strong pure-state stabilizer monotones for every integer $\alpha\ge 2$.\label{th:stabpuritystrongmonotones}
\end{theorem}
\noindent
{\em Proof sketch.} To show the strong pure-state monotonicity, we prove the more general result from which \cref{th:convexroofextension} follows. Starting from a pure state $\psi$, for all stabilizer operations $\mathcal{E}\in\mathcal{S}$ such that $\mathcal{E}(\psi)=\{(p_i,\rho_i)\}$, it holds that
\be M_{\alpha}^{\mathrm{lin}}(\psi)\ge \inf_{\substack{(q_{ij},\phi_{ij})}}\Big\{\sum_{ij}q_{ij}M_{\alpha}^{\mathrm{lin}}(\phi_{ij}): p_i\rho_i =\sum_jq_{ij} \phi_{ij}\Big\}\,.\nonumber
\ee
Similarly to \cref{th:stabentropypuremonotones}, the crucial step is to consider an arbitrary pure state $\ket{\psi}=\sqrt{p}\ket{0}\otimes\ket{\phi_1}+\sqrt{1-p}\ket{1}\otimes\ket{\phi_2}$ and show that $M_{\alpha}^{\mathrm{lin}}(\psi)\ge p M_{\alpha}^{\mathrm{lin}}(\phi_1)+ (1-p) M_{\alpha}^{\mathrm{lin}}(\phi_2)$. With the restriction imposed on the collection $\mathcal{E}(\psi)$ to consist solely of pure states, the result follows~\cite{_see_}.\qed

According to \cref{th:stabpuritystrongmonotones}, linear stabilizer entropies are  powerful stabilizer monotones defined through entropies of the characteristic distribution. In Ref.~\cite{haug_stabilizer_2023}, it was shown that stabilizer entropies $M_{\alpha}$ violate the strong monotonicity condition. However, since the two quantities are related to each other through a concave function, as $M_{\alpha}=\frac{1}{1-\alpha}\log_2 (1-M_{\alpha}^{\mathrm{lin}})$, ultimately this discrepancy is due to the Jensen's inequality which is equivalent to the difference between arithmetic mean and geometric mean. 
Let us present a counterexample for $M_{\alpha}$ for $\alpha\ge2$. We consider the state $\ket{\psi}\propto\ket{0}^{\otimes n}+\ket{1}\otimes\ket{\phi}$ with $\ket{\phi}$ a Haar random state on $n-1$ qubits. We have $M_{\alpha}(\ket{\psi})=O(1)$, as $\ket{\psi}$ has a non-vanishing overlap to the stabilizer state $\ket{0}^{\otimes n}$ (see~\cite{_see_}). While $M_{\alpha}(\ket{\phi})=\Omega(n)$ (with overwhelming probability~\cite{gu2023little}). As such, there exist $n_0$ such that $\forall n\ge n_0$ one has $M_{\alpha}(\ket{\psi})\le \frac{1}{2} M_{\alpha}(\ket{\phi})$, thus violating strong monotonicity of  $M_{\alpha}$ for some sufficiently large $n$. 

However, as a corollary of \cref{th:stabentropypuremonotones}, one can see that stabilizer entropies obey a weaker condition compared to strong pure-state monotonicity. Given a non-deterministic stabilizer protocol that, starting from a pure state $\ket{\psi}$, yields a collection of pure states $\mathcal{E}(\psi)=\{(p_i,\ket{\phi_i})\}$, it follows that $M_{\alpha}(\psi)\ge\min_i M_{\alpha}(\phi_i)$ for every integer $\alpha\ge 2$.

That being said, while strong monotonicity is indeed a desirable trait with clear operational significance, it is worth noting that it is not an essential requirement for effective stabilizer monotones. Min-relative entropy of magic~\cite{liu_manybody_2022}, which is a natural monotone being the (negative log) value of the overlap with the closest stabilizer state, does not obey strong monotonicity either (the counterexample shown above holds true). Consequently, the adherence of linear stabilizer entropies to strong monotonicity can be regarded as a distinctive and advantageous feature, {which we employ in the subsequent section to provide improved conversion bounds.}

{\em Improved resource conversion bounds.---} Having established the monotonicity of stabilizer entropy, we now exploit it to improve on resource conversion bounds previously studied in the literature. Specifically, we build on the literature aiming to answer the following question: What is the rate of conversion between resource states $\ket{R_1}$ and $\ket{R_2}$, that is, the maximum number of copies of a given resource state $\ket{R_2}$ produced per copy of another resource state $\ket{R_1}$ by means of stabilizer operations? 

Given a pure-state monotone $\mathcal{M}$, satisfying the additivity property, i.e. $\mathcal{M}(\ket{\phi_1}\otimes\ket{\phi_2})=\mathcal{M}(\ket{\phi_1})+ \mathcal{M}(\ket{\phi_2})$, the rate of conversion $r_{\rightarrow}\coloneqq r[\ket{R_1}\mapsto \ket{R_2}]$ can be simply upper bounded by~\cite{beverland_lower_2020}
\be
r_{\rightarrow}\le\frac{\mathcal{M}(\ket{R_1})}{\mathcal{M}(\ket{R_2})}\,.
\ee
While one may expect that the rate $r_{\rightarrow}$ is simply the inverse of the rate $r_{\leftarrow}\coloneqq r[\ket{R_2}\mapsto \ket{R_1}]$, it has been shown that, in magic-state resource theory, there is a clear gap between $r_{\rightarrow}$ and $r_{\leftarrow}$~\cite{beverland_lower_2020}. In what follows, we build on this picture by analyzing conversion rates between the $m$-qubit resource states $\ket{C^{m-1}Z}\propto \sum_{b}(-1)^{b_1\cdots b_m}\ket{b}$ for $b\in\{0,1\}^m$. In particular, we analyze the gap between the forward conversion rate $r_{\rightarrow}^{(m)} \coloneqq r[\ket{C^{m-1}Z} \mapsto \ket{C^2Z}]$ and the backward conversion rate $r^{(m)}{\leftarrow} \coloneqq r[\ket{C^2Z} \mapsto \ket{C^{m-1}Z}]$ for $m \ge 3$. In Ref.~\cite{beverland_lower_2020}, via stabilizer nullity arguments, it has been shown that, to deterministically distill $\ket{C^{m-1}Z}$, it is not possible to consume fewer than $m/3$ $\ket{C^2Z}$, i.e., $r_{\leftarrow}^{(m)} \ge m/3$. We now investigate the forward conversion rate $r_{\rightarrow}^{(m)}$. The best known bound is $r_{\rightarrow}^{(4)} \le 4/3$. While already showing a clear gap, i.e. $\Omega(m)$ versus $O(1)$, between forward and backward rates, this bound leaves open the possibility of a forward conversion rate greater than one. The following corollary not only bounds $r_{\rightarrow}^{(4)}$ away from one, but also drastically amplifies the gap between forward and backward conversion rates.
\begin{corollary}
The conversion rate $r_{\rightarrow}^{(m)}\coloneqq r[\ket{C^{m-1}Z}\mapsto \ket{C^2Z}]=O(2^{-m})$. In particular, $r^{(m)}_{\rightarrow}< r^{(4)}_{\rightarrow}\le 0.9$ for every $m\in\mathbb{N}$.
\begin{proof}
It is sufficient to compute the $\alpha$-stabilizer entropy of $\ket{C^{m-1}Z}$ for $m\ge 3$, which can be simply done by a direct calculation. See Ref.~\cite{_see_} for the complete expression. Thanks to the additivity property of $M_{\alpha}$, we bound $r_{\rightarrow}^{(m)}\le M_{\alpha}(\ket{C^{m-1}Z})/M_{\alpha}(\ket{C^2Z}),\,\forall\alpha\in\mathbb{N}$~\cite{beverland_lower_2020}. The tightest bound is obtained for $\alpha=2$. 
\end{proof}
\end{corollary}
While the above bounds show an exponentially large gap, i.e., $\Omega(m)$ versus $O(2^{-m})$, between the forward and backward conversion rates, they focus on the \textit{deterministic} conversion between resource states. One may consider probabilistic protocols between resource states and ask whether one can achieve better conversion rates. A figure of merit for probabilistic conversion is given by  the optimal success probability $\pi^{\max}_{\ket{R_1}}(\ket{R_2})$ of converting the resource state $\ket{R_1}$ into $\ket{R_2}$. The latter can be nicely upper bounded using any strong stabilizer monotone $\mathcal{M}$ (see \cref{def:strongmagicmonotone}) as~\cite{chitambar_quantum_2019}
\be
\pi^{\max}_{\ket{R_1}}(\ket{R_2})\le \frac{\mathcal{M}(\ket{R_1})}{\mathcal{M}(\ket{R_2})}\,.\label{eqstrongprobability}
\ee
Hence, as a direct corollary of \cref{th:stabpuritystrongmonotones}, we answer the above question in the negative by showing an exponentially suppressed success probability. 
\begin{corollary}
The maximal probability for the probabilistic resource conversion $\ket{C^{m-1}Z}\mapsto \ket{C^2Z}$ obeys $\pi^{\max}_{\ket{C^{m-1}Z}}(\ket{C^2Z})=O(2^{-m})$.
\begin{proof}
    Computing the stabilizer purity of $\ket{C^{m-1}Z}$ for $m\ge 3$~\cite{_see_}, it is sufficient to exploit the strong monotonicity of linear stabilizer entropies and use Eq.~\eqref{eqstrongprobability}.
\end{proof}
\end{corollary}
These results, among others presented in Ref.~\cite{_see_}, strengthen the main findings of Ref.~\cite{beverland_lower_2020}: we can conclude that, even for probabilistic protocols, there is a clear preferred direction of conversion in magic-state resource theory.

{\em Extension to mixed states.---} While stabilizer entropies find their natural definition within the realm of pure states, when extending magic-state resource theory to mixed states, a diverse landscape emerges. This becomes evident when considering that magic-state auxiliary qubits are often assumed to be noisy, prompting interest in extracting a single clean non-Clifford unitary from many copies of these noisy auxiliary systems~\cite{bravyi_universal_2005}. To conclude, we leverage our findings to extend stabilizer entropies to mixed states, {via convex-roof extension~\cite{chitambar_quantum_2019}}, ensuring adherence to the monotonicity criterion outlined in \cref{def:magicmonotone}.
\begin{definition}[Convex-roof extension of SEs]\label{def:convexroof} Let $\mathcal{C}\coloneqq\{(p_i,\rho_i)\}$ be a collection of $n_i$ qubit quantum states. Extended stabilizer entropies are defined as
\begin{align}
    \widehat{M}_{\alpha}(\mathcal{C})&\coloneqq \frac{1}{1-\alpha}\log_2 \widehat{P}_{\alpha}(\mathcal{C})\,,\,\,\,\,\,\widehat{M}_{\alpha}^{\mathrm{lin}}(\mathcal{C})\coloneqq1-\widehat{P}_{\alpha}(\mathcal{C})
    \\
    \widehat{P}_{\alpha}(\mathcal{C})&\coloneqq\!\!\!\sup_{(q_{ij},\phi_{ij})}\!\Big\{\sum_{ij}q_{ij}P_{\alpha}(\phi_{ij})\Big\}\nonumber\,.
\end{align}
The $\sup$ is over the possible convex pure-state decompositions $p_i\rho_i=\sum_jq_{ij}\phi_{ij}$ for $\{(p_i,\rho_i)\}\in\mathcal{C}$. 
\end{definition}
It is worth noting that both the \textit{extended} stabilizer entropy and linear stabilizer entropy reduce to the entropies defined on pure states in \cref{def:stabilizerentropy}. The following theorem establishes the monotonicity of the above convex roof constructions.

\begin{theorem}\label{th:convexroofextension} Extended stabilizer entropies $\widehat{M}_{\alpha}$ and linear stabilizer entropies $\widehat{M}_{\alpha}^{\mathrm{lin}}$ are stabilizer monotones for every integer $\alpha\ge 2$. Moreover, $\widehat{M}_{\alpha}^{\mathrm{lin}}$ is a strong stabilizer monotone.
\end{theorem}
\noindent
{\em Proof sketch.} Both results for extended stabilizer entropy and linear stabilizer entropy follows by showing that, given a collection $\mathcal{C}=\{(p_i,\rho_i)\}$ of $n_i$ qubit states and an elementary stabilizer protocol $\mathcal{E}\in\mathcal{S}$, then: 
\be\quad \widehat P_{\alpha}[\mathcal{E}(\mathcal{C})]\geq \widehat P_{\alpha}(\mathcal{C})\,.
\ee
The steps follow from the proof of \cref{th:stabpuritystrongmonotones}, see Ref.~\cite{_see_}.\qed

The above result extends stabilizer entropies, along with some of their amenable properties, to the realm of mixed states, thus unlocking the exploration of the magic-state resource theory in many-body physics in more intricate situations. {While finding the optimal decomposition can be challenging in the worst case, there are scenarios where computing the extended stabilizer entropy is significantly less expensive than optimizing over stabilizer states, as other known measures require~\cite{liu_manybody_2022}. Let us estimate the cost of computing $\widehat{M}_{\alpha}$ for a $d^n$-dimensional density matrix of rank $r$. Using a brute-force search algorithm over convex decompositions, the computational effort involves $r4^{n}2^{O(nr^2)}$ time steps. This cost includes the $4^{n}$ operations needed to compute the stabilizer entropy $M_{\alpha}$ for $r$ pure states and the $2^{O(nr^2)}$ possible convex decompositions of $\rho$, resolving the stabilizer purity with $O(2^{-n})$ accuracy. Hence, for low-rank density matrices $r=o(\sqrt{n})$, this approach attains a super-polynomial advantage over the $2^{\Theta(n^2)}$ required to optimize over stabilizer decompositions.}


{\em Discussion.---} In this work, we prove that stabilizer entropies serve as  monotones for magic-state resource theory. Furthermore, their linearized counterparts exhibit strong monotonicity, meaning they do not, on average, increase following a non-deterministic protocol. {As a direct corollary of these findings, we provide enhanced conversion bounds between resource states. Aligning with the recent findings of Ref.~\cite{beverland_lower_2020}, our results indicate a preferred direction of conversion between magic states. }

{Additionally, as a result of our investigation, we provide a convex-roof extension of stabilizer entropies to mixed states, ensuring monotonicity under stabilizer protocols and achieving computational advantages over stabilizer decomposition for low-rank density matrices. This paves the way for further exploration of magic-state resource theory beyond pure states.}

While our results solidify the utility of stabilizer entropies in quantifying ``magic'' within multiqubit systems, there remain intriguing open questions regarding their properties, which remain mostly unexplored. For instance, despite robust metrics such as robustness of magic, stabilizer extent, and stabilizer nullity directly quantifying the cost of classical simulations via stabilizer formalism, there lacks a simulation algorithm directly assessed by stabilizer entropy {(with $\alpha\ge 2$)} costs.  {In this context, it is worth noting that the special case of $\alpha=1/2$ stabilizer entropy, also known as GKP magic~\cite{hahn_quantifying_2022} or stabilizer norm~\cite{howard_application_2017}, quantifies the simulation cost of classical simulation~\cite{PhysRevA.99.062337}, thus proving to be a powerful tool. However, it does not serve a monotone for magic-state resource theory~\cite{PhysRevLett.131.049901}.}

The maximal resourceful {$n$-qubit} state according to SEs remains unknown. While it is evident that the upper bound $M_{2}<\log_2\frac{d_n+1}{2}$ cannot be achieved by any characteristic distribution (for $n\neq1,3$), a tighter bound and the state saturating it are yet to be determined.   

In conclusion, this Letter provides a definitive answer to the open question of whether stabilizer entropies serve as monotones for magic-state resource theory or merely act as weak detectors of nonstabilizerness. With this clarification, there arises a compelling need for further exploration of stabilizer entropies across diverse domains, ranging from many-body physics to classical simulation via stabilizer formalism, and within the realms of theoretical resource theory.

{\em Acknowledgments.---} We acknowledge an inspiring discussion with Ludovico Lami. We are grateful to Salvatore F.E. Oliviero for the careful reading of the manuscript. We thank Alioscia Hamma and Francesco A. Mele for their valuable comments and discussion.  We thank Ryuji Takagi for having pointed out a mistake in an earlier version. L.L. is funded through the Munich Quantum Valley project (MQV-K8) by Bayerisches Staatsministerium für Wissenschaft und Kunst. L.B. was funded by  DFG (FOR 2724, CRC 183), the Cluster of Excellence MATH+ and the BMBF (MuniQC-Atoms).



\begin{thebibliography}{35}%
\makeatletter
\providecommand \@ifxundefined [1]{%
 \@ifx{#1\undefined}
}%
\providecommand \@ifnum [1]{%
 \ifnum #1\expandafter \@firstoftwo
 \else \expandafter \@secondoftwo
 \fi
}%
\providecommand \@ifx [1]{%
 \ifx #1\expandafter \@firstoftwo
 \else \expandafter \@secondoftwo
 \fi
}%
\providecommand \natexlab [1]{#1}%
\providecommand \enquote  [1]{``#1''}%
\providecommand \bibnamefont  [1]{#1}%
\providecommand \bibfnamefont [1]{#1}%
\providecommand \citenamefont [1]{#1}%
\providecommand \href@noop [0]{\@secondoftwo}%
\providecommand \href [0]{\begingroup \@sanitize@url \@href}%
\providecommand \@href[1]{\@@startlink{#1}\@@href}%
\providecommand \@@href[1]{\endgroup#1\@@endlink}%
\providecommand \@sanitize@url [0]{\catcode `\\12\catcode `\$12\catcode `\&12\catcode `\#12\catcode `\^12\catcode `\_12\catcode `\%12\relax}%
\providecommand \@@startlink[1]{}%
\providecommand \@@endlink[0]{}%
\providecommand \url  [0]{\begingroup\@sanitize@url \@url }%
\providecommand \@url [1]{\endgroup\@href {#1}{\urlprefix }}%
\providecommand \urlprefix  [0]{URL }%
\providecommand \Eprint [0]{\href }%
\providecommand \doibase [0]{http://dx.doi.org/}%
\providecommand \selectlanguage [0]{\@gobble}%
\providecommand \bibinfo  [0]{\@secondoftwo}%
\providecommand \bibfield  [0]{\@secondoftwo}%
\providecommand \translation [1]{[#1]}%
\providecommand \BibitemOpen [0]{}%
\providecommand \bibitemStop [0]{}%
\providecommand \bibitemNoStop [0]{.\EOS\space}%
\providecommand \EOS [0]{\spacefactor3000\relax}%
\providecommand \BibitemShut  [1]{\csname bibitem#1\endcsname}%
\let\auto@bib@innerbib\@empty
\bibitem [{\citenamefont {Bravyi}\ and\ \citenamefont {Kitaev}(2005)}]{bravyi_universal_2005}%
  \BibitemOpen
  \bibfield  {author} {\bibinfo {author} {\bibfnamefont {S.}~\bibnamefont {Bravyi}}\ and\ \bibinfo {author} {\bibfnamefont {A.}~\bibnamefont {Kitaev}},\ }\href {\doibase 10.1103/PhysRevA.71.022316} {\bibfield  {journal} {\bibinfo  {journal} {Physical Review A}\ }\textbf {\bibinfo {volume} {71}},\ \bibinfo {pages} {022316} (\bibinfo {year} {2005})}\BibitemShut {NoStop}%
\bibitem [{\citenamefont {Veitch}\ \emph {et~al.}(2014)\citenamefont {Veitch}, \citenamefont {Hamed~Mousavian}, \citenamefont {Gottesman},\ and\ \citenamefont {Emerson}}]{Veitch_2014}%
  \BibitemOpen
  \bibfield  {author} {\bibinfo {author} {\bibfnamefont {V.}~\bibnamefont {Veitch}}, \bibinfo {author} {\bibfnamefont {S.~A.}\ \bibnamefont {Hamed~Mousavian}}, \bibinfo {author} {\bibfnamefont {D.}~\bibnamefont {Gottesman}}, \ and\ \bibinfo {author} {\bibfnamefont {J.}~\bibnamefont {Emerson}},\ }\href {\doibase 10.1088/1367-2630/16/1/013009} {\bibfield  {journal} {\bibinfo  {journal} {New Journal of Physics}\ }\textbf {\bibinfo {volume} {16}},\ \bibinfo {pages} {013009} (\bibinfo {year} {2014})}\BibitemShut {NoStop}%
\bibitem [{\citenamefont {Campbell}\ and\ \citenamefont {Browne}(2010)}]{campbell_bound_2010}%
  \BibitemOpen
  \bibfield  {author} {\bibinfo {author} {\bibfnamefont {E.~T.}\ \bibnamefont {Campbell}}\ and\ \bibinfo {author} {\bibfnamefont {D.~E.}\ \bibnamefont {Browne}},\ }\href {\doibase 10.1103/PhysRevLett.104.030503} {\bibfield  {journal} {\bibinfo  {journal} {Physical Review Letters}\ }\textbf {\bibinfo {volume} {104}},\ \bibinfo {pages} {030503} (\bibinfo {year} {2010})}\BibitemShut {NoStop}%
\bibitem [{\citenamefont {Eastin}\ and\ \citenamefont {Knill}(2009)}]{PhysRevLett.102.110502}%
  \BibitemOpen
  \bibfield  {author} {\bibinfo {author} {\bibfnamefont {B.}~\bibnamefont {Eastin}}\ and\ \bibinfo {author} {\bibfnamefont {E.}~\bibnamefont {Knill}},\ }\href {\doibase 10.1103/PhysRevLett.102.110502} {\bibfield  {journal} {\bibinfo  {journal} {Phys. Rev. Lett.}\ }\textbf {\bibinfo {volume} {102}},\ \bibinfo {pages} {110502} (\bibinfo {year} {2009})}\BibitemShut {NoStop}%
\bibitem [{\citenamefont {Gottesman}(1998)}]{gottesman_heisenberg_1998}%
  \BibitemOpen
  \bibfield  {author} {\bibinfo {author} {\bibfnamefont {D.}~\bibnamefont {Gottesman}},\ }\href {\doibase 10.48550/arXiv.quant-ph/9807006} {\enquote {\bibinfo {title} {The {{Heisenberg Representation}} of {{Quantum Computers}}},}\ } (\bibinfo {year} {1998}),\ \Eprint {http://arxiv.org/abs/quant-ph/9807006} {arxiv:quant-ph/9807006} \BibitemShut {NoStop}%
\bibitem [{\citenamefont {Aaronson}\ and\ \citenamefont {Gottesman}(2004)}]{aaronson_improved_2004}%
  \BibitemOpen
  \bibfield  {author} {\bibinfo {author} {\bibfnamefont {S.}~\bibnamefont {Aaronson}}\ and\ \bibinfo {author} {\bibfnamefont {D.}~\bibnamefont {Gottesman}},\ }\href {\doibase 10.1103/PhysRevA.70.052328} {\bibfield  {journal} {\bibinfo  {journal} {Physical Review A}\ }\textbf {\bibinfo {volume} {70}},\ \bibinfo {pages} {052328} (\bibinfo {year} {2004})}\BibitemShut {NoStop}%
\bibitem [{\citenamefont {Chitambar}\ and\ \citenamefont {Gour}(2019)}]{chitambar_quantum_2019}%
  \BibitemOpen
  \bibfield  {author} {\bibinfo {author} {\bibfnamefont {E.}~\bibnamefont {Chitambar}}\ and\ \bibinfo {author} {\bibfnamefont {G.}~\bibnamefont {Gour}},\ }\href {\doibase 10.1103/RevModPhys.91.025001} {\bibfield  {journal} {\bibinfo  {journal} {Review of Modern Physics}\ }\textbf {\bibinfo {volume} {91}},\ \bibinfo {pages} {025001} (\bibinfo {year} {2019})}\BibitemShut {NoStop}%
\bibitem [{\citenamefont {Leone}\ \emph {et~al.}(2022)\citenamefont {Leone}, \citenamefont {Oliviero},\ and\ \citenamefont {Hamma}}]{leone_stabilizer_2022}%
  \BibitemOpen
  \bibfield  {author} {\bibinfo {author} {\bibfnamefont {L.}~\bibnamefont {Leone}}, \bibinfo {author} {\bibfnamefont {S.~F.~E.}\ \bibnamefont {Oliviero}}, \ and\ \bibinfo {author} {\bibfnamefont {A.}~\bibnamefont {Hamma}},\ }\href {\doibase 10.1103/PhysRevLett.128.050402} {\bibfield  {journal} {\bibinfo  {journal} {Physical Review Letters}\ }\textbf {\bibinfo {volume} {128}},\ \bibinfo {pages} {050402} (\bibinfo {year} {2022})}\BibitemShut {NoStop}%
\bibitem [{\citenamefont {Passarelli}\ \emph {et~al.}(2024)\citenamefont {Passarelli}, \citenamefont {Fazio},\ and\ \citenamefont {Lucignano}}]{passarelli2024nonstabilizerness}%
  \BibitemOpen
  \bibfield  {author} {\bibinfo {author} {\bibfnamefont {G.}~\bibnamefont {Passarelli}}, \bibinfo {author} {\bibfnamefont {R.}~\bibnamefont {Fazio}}, \ and\ \bibinfo {author} {\bibfnamefont {P.}~\bibnamefont {Lucignano}},\ }\href {\doibase 10.1103/PhysRevA.110.022436} {\bibfield  {journal} {\bibinfo  {journal} {Phys. Rev. A}\ }\textbf {\bibinfo {volume} {110}},\ \bibinfo {pages} {022436} (\bibinfo {year} {2024})}\BibitemShut {NoStop}%
\bibitem [{\citenamefont {Oliviero}\ \emph {et~al.}(2022{\natexlab{a}})\citenamefont {Oliviero}, \citenamefont {Leone},\ and\ \citenamefont {Hamma}}]{PhysRevA.106.042426}%
  \BibitemOpen
  \bibfield  {author} {\bibinfo {author} {\bibfnamefont {S.~F.~E.}\ \bibnamefont {Oliviero}}, \bibinfo {author} {\bibfnamefont {L.}~\bibnamefont {Leone}}, \ and\ \bibinfo {author} {\bibfnamefont {A.}~\bibnamefont {Hamma}},\ }\href {\doibase 10.1103/PhysRevA.106.042426} {\bibfield  {journal} {\bibinfo  {journal} {Phys. Rev. A}\ }\textbf {\bibinfo {volume} {106}},\ \bibinfo {pages} {042426} (\bibinfo {year} {2022}{\natexlab{a}})}\BibitemShut {NoStop}%
\bibitem [{\citenamefont {Haug}\ and\ \citenamefont {Piroli}(2023{\natexlab{a}})}]{haug_quantifying_2023}%
  \BibitemOpen
  \bibfield  {author} {\bibinfo {author} {\bibfnamefont {T.}~\bibnamefont {Haug}}\ and\ \bibinfo {author} {\bibfnamefont {L.}~\bibnamefont {Piroli}},\ }\href {\doibase 10.1103/PhysRevB.107.035148} {\bibfield  {journal} {\bibinfo  {journal} {Phys. Rev. B}\ }\textbf {\bibinfo {volume} {107}},\ \bibinfo {pages} {035148} (\bibinfo {year} {2023}{\natexlab{a}})}\BibitemShut {NoStop}%
\bibitem [{\citenamefont {Chen}\ \emph {et~al.}(2024)\citenamefont {Chen}, \citenamefont {Yan},\ and\ \citenamefont {Zhou}}]{chen2023magic2}%
  \BibitemOpen
  \bibfield  {author} {\bibinfo {author} {\bibfnamefont {J.}~\bibnamefont {Chen}}, \bibinfo {author} {\bibfnamefont {Y.}~\bibnamefont {Yan}}, \ and\ \bibinfo {author} {\bibfnamefont {Y.}~\bibnamefont {Zhou}},\ }\href {\doibase 10.22331/q-2024-05-21-1351} {\bibfield  {journal} {\bibinfo  {journal} {Quantum}\ }\textbf {\bibinfo {volume} {8}},\ \bibinfo {pages} {1351} (\bibinfo {year} {2024})}\BibitemShut {NoStop}%
\bibitem [{\citenamefont {Lami}\ and\ \citenamefont {Collura}(2023)}]{lami_quantum_2023}%
  \BibitemOpen
  \bibfield  {author} {\bibinfo {author} {\bibfnamefont {G.}~\bibnamefont {Lami}}\ and\ \bibinfo {author} {\bibfnamefont {M.}~\bibnamefont {Collura}},\ }\href {\doibase 10.1103/PhysRevLett.131.180401} {\bibfield  {journal} {\bibinfo  {journal} {Phys. Rev. Lett.}\ }\textbf {\bibinfo {volume} {131}},\ \bibinfo {pages} {180401} (\bibinfo {year} {2023})}\BibitemShut {NoStop}%
\bibitem [{\citenamefont {Tarabunga}\ \emph {et~al.}(2024)\citenamefont {Tarabunga}, \citenamefont {Tirrito}, \citenamefont {Bañuls},\ and\ \citenamefont {Dalmonte}}]{tarabunga2024nonstabilizerness}%
  \BibitemOpen
  \bibfield  {author} {\bibinfo {author} {\bibfnamefont {P.~S.}\ \bibnamefont {Tarabunga}}, \bibinfo {author} {\bibfnamefont {E.}~\bibnamefont {Tirrito}}, \bibinfo {author} {\bibfnamefont {M.~C.}\ \bibnamefont {Bañuls}}, \ and\ \bibinfo {author} {\bibfnamefont {M.}~\bibnamefont {Dalmonte}},\ }\href {\doibase 10.1103/physrevlett.133.010601} {\bibfield  {journal} {\bibinfo  {journal} {Physical Review Letters}\ }\textbf {\bibinfo {volume} {133}} (\bibinfo {year} {2024}),\ 10.1103/physrevlett.133.010601}\BibitemShut {NoStop}%
\bibitem [{\citenamefont {Oliviero}\ \emph {et~al.}(2022{\natexlab{b}})\citenamefont {Oliviero}, \citenamefont {Leone}, \citenamefont {Hamma},\ and\ \citenamefont {Lloyd}}]{oliviero2022MeasuringMagicQuantum}%
  \BibitemOpen
  \bibfield  {author} {\bibinfo {author} {\bibfnamefont {S.~F.~E.}\ \bibnamefont {Oliviero}}, \bibinfo {author} {\bibfnamefont {L.}~\bibnamefont {Leone}}, \bibinfo {author} {\bibfnamefont {A.}~\bibnamefont {Hamma}}, \ and\ \bibinfo {author} {\bibfnamefont {S.}~\bibnamefont {Lloyd}},\ }\href {\doibase 10.1038/s41534-022-00666-5} {\bibfield  {journal} {\bibinfo  {journal} {npj Quantum Inf}\ }\textbf {\bibinfo {volume} {8}},\ \bibinfo {pages} {1} (\bibinfo {year} {2022}{\natexlab{b}})}\BibitemShut {NoStop}%
\bibitem [{\citenamefont {Haug}\ \emph {et~al.}(2024)\citenamefont {Haug}, \citenamefont {Lee},\ and\ \citenamefont {Kim}}]{haug_efficient_2023}%
  \BibitemOpen
  \bibfield  {author} {\bibinfo {author} {\bibfnamefont {T.}~\bibnamefont {Haug}}, \bibinfo {author} {\bibfnamefont {S.}~\bibnamefont {Lee}}, \ and\ \bibinfo {author} {\bibfnamefont {M.~S.}\ \bibnamefont {Kim}},\ }\href {\doibase 10.1103/PhysRevLett.132.240602} {\bibfield  {journal} {\bibinfo  {journal} {Phys. Rev. Lett.}\ }\textbf {\bibinfo {volume} {132}},\ \bibinfo {pages} {240602} (\bibinfo {year} {2024})}\BibitemShut {NoStop}%
\bibitem [{\citenamefont {Heinrich}\ and\ \citenamefont {Gross}(pril)}]{heinrich_robustness_2019}%
  \BibitemOpen
  \bibfield  {author} {\bibinfo {author} {\bibfnamefont {M.}~\bibnamefont {Heinrich}}\ and\ \bibinfo {author} {\bibfnamefont {D.}~\bibnamefont {Gross}},\ }\href {\doibase 10.22331/q-2019-04-08-132} {\bibfield  {journal} {\bibinfo  {journal} {Quantum}\ }\textbf {\bibinfo {volume} {3}},\ \bibinfo {pages} {132} (\bibinfo {year} {2019/april})}\BibitemShut {NoStop}%
\bibitem [{\citenamefont {Gu}\ \emph {et~al.}(2024)\citenamefont {Gu}, \citenamefont {Leone}, \citenamefont {Ghosh}, \citenamefont {Eisert}, \citenamefont {Yelin},\ and\ \citenamefont {Quek}}]{gu2023little}%
  \BibitemOpen
  \bibfield  {author} {\bibinfo {author} {\bibfnamefont {A.}~\bibnamefont {Gu}}, \bibinfo {author} {\bibfnamefont {L.}~\bibnamefont {Leone}}, \bibinfo {author} {\bibfnamefont {S.}~\bibnamefont {Ghosh}}, \bibinfo {author} {\bibfnamefont {J.}~\bibnamefont {Eisert}}, \bibinfo {author} {\bibfnamefont {S.~F.}\ \bibnamefont {Yelin}}, \ and\ \bibinfo {author} {\bibfnamefont {Y.}~\bibnamefont {Quek}},\ }\href {\doibase 10.1103/physrevlett.132.210602} {\bibfield  {journal} {\bibinfo  {journal} {Physical Review Letters}\ }\textbf {\bibinfo {volume} {132}} (\bibinfo {year} {2024}),\ 10.1103/physrevlett.132.210602}\BibitemShut {NoStop}%
\bibitem [{\citenamefont {Leone}\ \emph {et~al.}(2023)\citenamefont {Leone}, \citenamefont {Oliviero},\ and\ \citenamefont {Hamma}}]{leone_nonstabilizerness_2023}%
  \BibitemOpen
  \bibfield  {author} {\bibinfo {author} {\bibfnamefont {L.}~\bibnamefont {Leone}}, \bibinfo {author} {\bibfnamefont {S.~F.~E.}\ \bibnamefont {Oliviero}}, \ and\ \bibinfo {author} {\bibfnamefont {A.}~\bibnamefont {Hamma}},\ }\href {\doibase 10.1103/PhysRevA.107.022429} {\bibfield  {journal} {\bibinfo  {journal} {Phys. Rev. A}\ }\textbf {\bibinfo {volume} {107}},\ \bibinfo {pages} {022429} (\bibinfo {year} {2023})}\BibitemShut {NoStop}%
\bibitem [{\citenamefont {Hinsche}\ \emph {et~al.}(2024)\citenamefont {Hinsche}, \citenamefont {Ioannou}, \citenamefont {Jerbi}, \citenamefont {Leone}, \citenamefont {Eisert},\ and\ \citenamefont {Carrasco}}]{hinsche2024efficient}%
  \BibitemOpen
  \bibfield  {author} {\bibinfo {author} {\bibfnamefont {M.}~\bibnamefont {Hinsche}}, \bibinfo {author} {\bibfnamefont {M.}~\bibnamefont {Ioannou}}, \bibinfo {author} {\bibfnamefont {S.}~\bibnamefont {Jerbi}}, \bibinfo {author} {\bibfnamefont {L.}~\bibnamefont {Leone}}, \bibinfo {author} {\bibfnamefont {J.}~\bibnamefont {Eisert}}, \ and\ \bibinfo {author} {\bibfnamefont {J.}~\bibnamefont {Carrasco}},\ }\href {https://arxiv.org/abs/2405.06544} {\enquote {\bibinfo {title} {Efficient distributed inner product estimation via pauli sampling},}\ } (\bibinfo {year} {2024}),\ \Eprint {http://arxiv.org/abs/2405.06544} {arXiv:2405.06544 [quant-ph]} \BibitemShut {NoStop}%
\bibitem [{\citenamefont {Leone}\ \emph {et~al.}(2024)\citenamefont {Leone}, \citenamefont {Oliviero},\ and\ \citenamefont {Hamma}}]{leone2023learning}%
  \BibitemOpen
  \bibfield  {author} {\bibinfo {author} {\bibfnamefont {L.}~\bibnamefont {Leone}}, \bibinfo {author} {\bibfnamefont {S.~F.~E.}\ \bibnamefont {Oliviero}}, \ and\ \bibinfo {author} {\bibfnamefont {A.}~\bibnamefont {Hamma}},\ }\href {\doibase 10.22331/q-2024-05-27-1361} {\bibfield  {journal} {\bibinfo  {journal} {Quantum}\ }\textbf {\bibinfo {volume} {8}},\ \bibinfo {pages} {1361} (\bibinfo {year} {2024})}\BibitemShut {NoStop}%
\bibitem [{\citenamefont {Zhang}\ \emph {et~al.}(2024)\citenamefont {Zhang}, \citenamefont {Liu},\ and\ \citenamefont {Zhou}}]{zhang2024minimal}%
  \BibitemOpen
  \bibfield  {author} {\bibinfo {author} {\bibfnamefont {Q.}~\bibnamefont {Zhang}}, \bibinfo {author} {\bibfnamefont {Q.}~\bibnamefont {Liu}}, \ and\ \bibinfo {author} {\bibfnamefont {Y.}~\bibnamefont {Zhou}},\ }\href {\doibase 10.1103/PhysRevApplied.21.064001} {\bibfield  {journal} {\bibinfo  {journal} {Phys. Rev. Appl.}\ }\textbf {\bibinfo {volume} {21}},\ \bibinfo {pages} {064001} (\bibinfo {year} {2024})}\BibitemShut {NoStop}%
\bibitem [{\citenamefont {Brieger}\ \emph {et~al.}(2023)\citenamefont {Brieger}, \citenamefont {Heinrich}, \citenamefont {Roth},\ and\ \citenamefont {Kliesch}}]{brieger2023stability}%
  \BibitemOpen
  \bibfield  {author} {\bibinfo {author} {\bibfnamefont {R.}~\bibnamefont {Brieger}}, \bibinfo {author} {\bibfnamefont {M.}~\bibnamefont {Heinrich}}, \bibinfo {author} {\bibfnamefont {I.}~\bibnamefont {Roth}}, \ and\ \bibinfo {author} {\bibfnamefont {M.}~\bibnamefont {Kliesch}},\ }\href@noop {} {\enquote {\bibinfo {title} {Stability of classical shadows under gate-dependent noise},}\ } (\bibinfo {year} {2023}),\ \Eprint {http://arxiv.org/abs/2310.19947} {arXiv:2310.19947 [quant-ph]} \BibitemShut {NoStop}%
\bibitem [{\citenamefont {Haug}\ and\ \citenamefont {Piroli}(2023{\natexlab{b}})}]{haug_stabilizer_2023}%
  \BibitemOpen
  \bibfield  {author} {\bibinfo {author} {\bibfnamefont {T.}~\bibnamefont {Haug}}\ and\ \bibinfo {author} {\bibfnamefont {L.}~\bibnamefont {Piroli}},\ }\href {\doibase 10.22331/q-2023-08-28-1092} {\bibfield  {journal} {\bibinfo  {journal} {Quantum}\ }\textbf {\bibinfo {volume} {7}},\ \bibinfo {pages} {1092} (\bibinfo {year} {2023}{\natexlab{b}})}\BibitemShut {NoStop}%
\bibitem [{\citenamefont {Howard}\ and\ \citenamefont {Campbell}(2017)}]{howard_application_2017}%
  \BibitemOpen
  \bibfield  {author} {\bibinfo {author} {\bibfnamefont {M.}~\bibnamefont {Howard}}\ and\ \bibinfo {author} {\bibfnamefont {E.}~\bibnamefont {Campbell}},\ }\href {\doibase 10.1103/PhysRevLett.118.090501} {\bibfield  {journal} {\bibinfo  {journal} {Physical Review Letters}\ }\textbf {\bibinfo {volume} {118}},\ \bibinfo {pages} {090501} (\bibinfo {year} {2017})}\BibitemShut {NoStop}%
\bibitem [{\citenamefont {Bravyi}\ and\ \citenamefont {Gosset}(2016)}]{bravyi_improved_2016}%
  \BibitemOpen
  \bibfield  {author} {\bibinfo {author} {\bibfnamefont {S.}~\bibnamefont {Bravyi}}\ and\ \bibinfo {author} {\bibfnamefont {D.}~\bibnamefont {Gosset}},\ }\href {\doibase 10.1103/PhysRevLett.116.250501} {\bibfield  {journal} {\bibinfo  {journal} {Physical Review Letters}\ }\textbf {\bibinfo {volume} {116}},\ \bibinfo {pages} {250501} (\bibinfo {year} {2016})}\BibitemShut {NoStop}%
\bibitem [{\citenamefont {Liu}\ and\ \citenamefont {Winter}(2022)}]{liu_manybody_2022}%
  \BibitemOpen
  \bibfield  {author} {\bibinfo {author} {\bibfnamefont {Z.-W.}\ \bibnamefont {Liu}}\ and\ \bibinfo {author} {\bibfnamefont {A.}~\bibnamefont {Winter}},\ }\href {\doibase 10.1103/PRXQuantum.3.020333} {\bibfield  {journal} {\bibinfo  {journal} {PRX Quantum}\ }\textbf {\bibinfo {volume} {3}},\ \bibinfo {pages} {020333} (\bibinfo {year} {2022})}\BibitemShut {NoStop}%
\bibitem [{\citenamefont {Beverland}\ \emph {et~al.}(2020)\citenamefont {Beverland}, \citenamefont {Campbell}, \citenamefont {Howard},\ and\ \citenamefont {Kliuchnikov}}]{beverland_lower_2020}%
  \BibitemOpen
  \bibfield  {author} {\bibinfo {author} {\bibfnamefont {M.}~\bibnamefont {Beverland}}, \bibinfo {author} {\bibfnamefont {E.}~\bibnamefont {Campbell}}, \bibinfo {author} {\bibfnamefont {M.}~\bibnamefont {Howard}}, \ and\ \bibinfo {author} {\bibfnamefont {V.}~\bibnamefont {Kliuchnikov}},\ }\href {\doibase 10.1088/2058-9565/ab8963} {\bibfield  {journal} {\bibinfo  {journal} {Quantum Science and Technology}\ }\textbf {\bibinfo {volume} {5}},\ \bibinfo {pages} {035009} (\bibinfo {year} {2020})}\BibitemShut {NoStop}%
\bibitem [{\citenamefont {Zhou}\ \emph {et~al.}(2000)\citenamefont {Zhou}, \citenamefont {Leung},\ and\ \citenamefont {Chuang}}]{Zhou_2000}%
  \BibitemOpen
  \bibfield  {author} {\bibinfo {author} {\bibfnamefont {X.}~\bibnamefont {Zhou}}, \bibinfo {author} {\bibfnamefont {D.~W.}\ \bibnamefont {Leung}}, \ and\ \bibinfo {author} {\bibfnamefont {I.~L.}\ \bibnamefont {Chuang}},\ }\href {\doibase 10.1103/physreva.62.052316} {\bibfield  {journal} {\bibinfo  {journal} {Physical Review A}\ }\textbf {\bibinfo {volume} {62}} (\bibinfo {year} {2000}),\ 10.1103/physreva.62.052316}\BibitemShut {NoStop}%
\bibitem [{\citenamefont {Zhu}\ \emph {et~al.}(2016)\citenamefont {Zhu}, \citenamefont {Kueng}, \citenamefont {Grassl},\ and\ \citenamefont {Gross}}]{zhu_clifford_2016}%
  \BibitemOpen
  \bibfield  {author} {\bibinfo {author} {\bibfnamefont {H.}~\bibnamefont {Zhu}}, \bibinfo {author} {\bibfnamefont {R.}~\bibnamefont {Kueng}}, \bibinfo {author} {\bibfnamefont {M.}~\bibnamefont {Grassl}}, \ and\ \bibinfo {author} {\bibfnamefont {D.}~\bibnamefont {Gross}},\ }\href {\doibase 10.48550/arXiv.1609.08172} {\enquote {\bibinfo {title} {{The {{Clifford}} Group Fails Gracefully to Be a Unitary 4-Design}},}\ } (\bibinfo {year} {2016}),\ \Eprint {http://arxiv.org/abs/1609.08172} {1609.08172 [quant-ph]} \BibitemShut {NoStop}%
\bibitem [{\citenamefont {Bravyi}\ \emph {et~al.}(2019)\citenamefont {Bravyi}, \citenamefont {Browne}, \citenamefont {Calpin}, \citenamefont {Campbell}, \citenamefont {Gosset},\ and\ \citenamefont {Howard}}]{Bravyi_2019}%
  \BibitemOpen
  \bibfield  {author} {\bibinfo {author} {\bibfnamefont {S.}~\bibnamefont {Bravyi}}, \bibinfo {author} {\bibfnamefont {D.}~\bibnamefont {Browne}}, \bibinfo {author} {\bibfnamefont {P.}~\bibnamefont {Calpin}}, \bibinfo {author} {\bibfnamefont {E.}~\bibnamefont {Campbell}}, \bibinfo {author} {\bibfnamefont {D.}~\bibnamefont {Gosset}}, \ and\ \bibinfo {author} {\bibfnamefont {M.}~\bibnamefont {Howard}},\ }\href {\doibase 10.22331/q-2019-09-02-181} {\bibfield  {journal} {\bibinfo  {journal} {Quantum}\ }\textbf {\bibinfo {volume} {3}},\ \bibinfo {pages} {181} (\bibinfo {year} {2019})}\BibitemShut {NoStop}%
\bibitem [{_se()}]{_see_}%
  \BibitemOpen
  \href@noop {} { {\bibinfo {title} {See supplemental material, for formal proofs.}}\ }\BibitemShut {NoStop}%
\bibitem [{\citenamefont {Hahn}\ \emph {et~al.}(2022)\citenamefont {Hahn}, \citenamefont {Ferraro}, \citenamefont {Hultquist}, \citenamefont {Ferrini} \emph {et~al.}}]{hahn_quantifying_2022}%
  \BibitemOpen
  \bibfield  {author} {\bibinfo {author} {\bibfnamefont {O.}~\bibnamefont {Hahn}}, \bibinfo {author} {\bibfnamefont {A.}~\bibnamefont {Ferraro}}, \bibinfo {author} {\bibfnamefont {L.}~\bibnamefont {Hultquist}}, \bibinfo {author} {\bibfnamefont {G.}~\bibnamefont {Ferrini}},  \emph {et~al.},\ }\href {\doibase 10.1103/PhysRevLett.128.210502} {\bibfield  {journal} {\bibinfo  {journal} {Physical Review Letters}\ }\textbf {\bibinfo {volume} {128}},\ \bibinfo {pages} {210502} (\bibinfo {year} {2022})}\BibitemShut {NoStop}%
\bibitem [{\citenamefont {Rall}\ \emph {et~al.}(2019)\citenamefont {Rall}, \citenamefont {Liang}, \citenamefont {Cook},\ and\ \citenamefont {Kretschmer}}]{PhysRevA.99.062337}%
  \BibitemOpen
  \bibfield  {author} {\bibinfo {author} {\bibfnamefont {P.}~\bibnamefont {Rall}}, \bibinfo {author} {\bibfnamefont {D.}~\bibnamefont {Liang}}, \bibinfo {author} {\bibfnamefont {J.}~\bibnamefont {Cook}}, \ and\ \bibinfo {author} {\bibfnamefont {W.}~\bibnamefont {Kretschmer}},\ }\href {\doibase 10.1103/PhysRevA.99.062337} {\bibfield  {journal} {\bibinfo  {journal} {Phys. Rev. A}\ }\textbf {\bibinfo {volume} {99}},\ \bibinfo {pages} {062337} (\bibinfo {year} {2019})}\BibitemShut {NoStop}%
\bibitem [{\citenamefont {Hahn}\ \emph {et~al.}(2023)\citenamefont {Hahn}, \citenamefont {Ferraro}, \citenamefont {Hultquist}, \citenamefont {Ferrini},\ and\ \citenamefont {Garc\'{\i}a-\'Alvarez}}]{PhysRevLett.131.049901}%
  \BibitemOpen
  \bibfield  {author} {\bibinfo {author} {\bibfnamefont {O.}~\bibnamefont {Hahn}}, \bibinfo {author} {\bibfnamefont {A.}~\bibnamefont {Ferraro}}, \bibinfo {author} {\bibfnamefont {L.}~\bibnamefont {Hultquist}}, \bibinfo {author} {\bibfnamefont {G.}~\bibnamefont {Ferrini}}, \ and\ \bibinfo {author} {\bibfnamefont {L.}~\bibnamefont {Garc\'{\i}a-\'Alvarez}},\ }\href {\doibase 10.1103/PhysRevLett.131.049901} {\bibfield  {journal} {\bibinfo  {journal} {Phys. Rev. Lett.}\ }\textbf {\bibinfo {volume} {131}},\ \bibinfo {pages} {049901} (\bibinfo {year} {2023})}\BibitemShut {NoStop}%
\end{thebibliography}

%

\appendix
\onecolumngrid

\section{Stabilizer entropy}\label{app:stabentropy}
In this section, we list the properties of stabilizer entropies explored in a number of works. First of all, let us recall the definition. Let $\psi$ be a pure state. Given the characteristic distribution $d^{-1}_n\tr^2(P\psi)$, stabilizer entropies are defined 
\be
M_{\alpha}(\psi)=\frac{1}{1-\alpha}\log \frac{1}{d_n}\sum_{P\in\mathbb{P}_n}\tr^{2\alpha}(P\psi)\,.
\ee
It obeys the following properties:
\begin{itemize}
    \item $M_{\alpha}(\sigma)=0$ if and only if $\sigma$ is a pure stabilizer state;
    \item  it is invariant under unitary Clifford operations $C$, i.e. $M_{\alpha}(C\psi C^{\dag})=M_{\alpha}(\psi)$;
    \item it is additive $M_{\alpha}(\psi\otimes \phi)=M_{\alpha}(\psi)+M_{\alpha}(\phi)$;
    \item ordered $M_{\alpha}(\psi)\ge M_{\beta}(\psi)$ if $\alpha<\beta$;
    \item it is bounded $M_{\alpha}\le \log d_n$ and, for every $\alpha\ge 2$, the tighter bound $M_{\alpha}\le \log (d_n+1)-1$ holds
    \item it lower bounds the stabilizer nullity $\nu$~\cite{beverland_lower_2020}, $M_{\alpha}(\psi)\le \nu(\psi)$. Given a state $\ket{\psi}$, the stabilizer nullity is proportional to the dimension $\nu(\psi)=n-\dim(G)$ of the largest Abelian subgroup $G$ of the Pauli group that stabilizes $\ket{\psi}$, i.e., $\forall P\in G$ then $P\ket{\psi}=\ket{\psi}$.
    \item for $\alpha\ge 2$, it lower bounds the min-relative entropy of magic~\cite{liu_manybody_2022}, i.e. $M_{\alpha}(\psi)\le \frac{2\alpha}{\alpha-1}D_{\min}(\psi)\coloneqq -\frac{2\alpha}{\alpha-1}\log\max_{\sigma}|\langle\psi|\sigma\rangle|^2$~\cite{haug_stabilizer_2023};
    \item  for $\alpha\ge\frac{1}{2}$, it lower bounds the Robustness of magic~\cite{leone_stabilizer_2022}, i.e. $M_{\alpha}\le 2\log\mathcal{R}(\psi)\coloneqq2\min_x\{\sum_i|x_i|\,:\, \rho=\sum_ix_i\sigma_i\,,\, \sigma_i\in\mathrm{STAB}\}$~\cite{howard_application_2017};
    \item  for $\alpha\ge 2$, it lower bounds the stabilizer extent~\cite{bravyi_improved_2016}, i.e. $M_{\alpha}\le 4\log \xi(\psi)\coloneqq\min\{\sum_i|x_i|\,:\, \ket{\psi}=\sum_ix_i\ket{\sigma_i}\,,\, \ket{\sigma_i}\in\mathrm{STAB}\}$;
    \item for $\alpha=1$, it obeys a Fannes-type inequality~\cite{gu2023little}. For any two states $\ket{\psi}$ and $\ket{\phi}$, it holds that
\begin{equation}\label{eq:Fannes}
\abs{M_{1}(\psi)-M_{1}(\phi)}\le\begin{cases} \norm{\psi-\phi}_1\log(d_n^2-1)+H_{\textrm{bin}}[\norm{\psi-\phi}_1] &\text{for $\norm{\psi-\phi}_1\le 1/2$},\\
\norm{\psi-\phi}_1\log(d_n^2-1)+1 &\text{for $\norm{\psi-\phi}_1> 1/2$}\\
\end{cases}
\end{equation}
\item for every odd integer $\alpha\ge 2$, it can be measured efficiently. In particular $P_{\alpha}(\psi)$ (stabilizer purity) can be measured using $O(n\epsilon^{-2})$ many samples of the state $\ket{\psi}$ and consuming two copies of the state at a time~\cite{haug_efficient_2023}.
\end{itemize}
. 
\section{Preliminary results}

\begin{lemma}\label{app:lemma2}
Consider $\ket{\psi}=\sqrt{p}\ket{0}\otimes\ket{\phi_1}+\sqrt{1-p}\ket{1}\otimes \ket{\phi_2}$ and $\ket{\phi_1},\ket{\phi_2}\in \mathbb{C}^{2\otimes (n-1)}$. For any integer $\alpha\ge 2$, it holds that
\be
P_{\alpha}(\psi)\le \sum_{i=0}^{\alpha}\binom{2\alpha}{2i}p^{2i}(1-p)^{2(\alpha-i)}P_{\alpha}^{\frac{i}{\alpha}}(\phi_1)P_{\alpha}^{\frac{\alpha-i}{\alpha}}(\phi_2)+2^{2\alpha-1}p^{\alpha}(1-p)^{\alpha}\sqrt{P_{\alpha}(\phi_1)P_{\alpha}(\phi_2)}\label{app:firstexpressionlemma2}\,.
\ee
Alternatively, the sum can be performed to obtain
\be
P_{\alpha}(\psi)\le \sum_{\pm}\frac{1}{2}\left(pP^{1/2\alpha}_\alpha(\phi_1)\pm(1-p)P^{1/2\alpha}_\alpha(\phi_2)\right)^{2\alpha}+2^{2\alpha-1}p^{\alpha}(1-p)^{\alpha}\sqrt{P_{\alpha}(\phi_1)P_{\alpha}(\phi_2)}\,.\label{app:secondexpressionlemma2}
\ee
\begin{proof}
First of all, we can express $P_{\alpha}(\psi)$ as
\ba
P_{\alpha}(\psi)&=&\frac{1}{d_n}\sum_{P}\tr^{2\alpha}(I\otimes P\psi)+\tr^{2\alpha}(Z\otimes P\psi)+\tr^{2\alpha}(X\otimes P\psi)+\tr^{2\alpha}(Y\otimes P\psi)\\
&=&\frac{1}{d_n}\sum_{P}\left(p\tr(P\phi_1)+(1-p)\tr(P\phi_2)\right)^{2\alpha}+\frac{1}{d_n}\sum_{P}\left(p\tr(P\phi_1)-(1-p)\tr(P\phi_2)\right)^{2\alpha}\label{A2}\\
&+&\frac{1}{d_n}\sum_P(\mathrm{Re}[2\sqrt{p(1-p)}\langle \phi_2|P|\phi_1\rangle])^{2\alpha}+\frac{1}{d_n}\sum_P(\mathrm{Im}[2\sqrt{p(1-p)}\langle \phi_2|P|\phi_1\rangle])^{2\alpha}\,,\label{A3}
\ea
where the sum runs over the Pauli group on $n-1$ qubits and $d_n\coloneqq2^n$. Let us first elaborate on \eqref{A2} and bound it as
\ba
&&\sum_{P}\left(p\tr(P\phi_1)+(1-p)\tr(P\phi_2)\right)^{2\alpha}+\sum_{P}\left(p\tr(P\phi_1)-(1-p)\tr(P\phi_2)\right)^{2\alpha}\\
&=&2\sum_P\sum_{i=0}^{\alpha}\binom{2\alpha}{2i}(p\tr(P\phi_1))^{2i}((1-p)\tr(P\phi_2))^{2(\alpha-i)}\nonumber\\
&\le& d_n\sum_{i=0}^{\alpha}\binom{2\alpha}{2i}p^{2i}(1-p)^{2(\alpha-i)}P_{\alpha}^{\frac{i}{\alpha}}(\phi_1)P_{\alpha}^{\frac{\alpha-i}{\alpha}}(\phi_2)\,.\label{a13}
\ea
In Eq.~\eqref{a13} we used Hölder's inequality $\sum_{l}x_{l}^{i}y_{l}^{\alpha-i}\le \|x\|_{\alpha/i}\|y\|_{\alpha/(\alpha-i)}$ with $i/\alpha+(\alpha-i)/\alpha=1$. Moreover, we used the definition of stabilizer purity $P_{\alpha}(\phi_i)=\frac{1}{d_{n-1}}\sum_P\tr^{2\alpha}(P\phi_i)$. Let us now analyze the term in Eq.~\eqref{A3}:
\ba
&&\sum_P(\mathrm{Re}[2\sqrt{p(1-p)}\langle \phi_2|P|\phi_1\rangle])^{2\alpha}+\sum_P(\mathrm{Im}[2\sqrt{p(1-p)}\langle \phi_2|P|\phi_1\rangle])^{2\alpha}
\nonumber\\&\le&2^{2\alpha}p^{\alpha}(1-p)^{\alpha}\sum_{P}\tr^{\alpha}(P\phi_1P\phi_2)\label{a14}\\
&\le&2^{2\alpha}p^{\alpha}(1-p)^{\alpha}\sqrt{\sum_{P}\tr^{2\alpha}(P\phi_1)}\sqrt{\sum_P\tr^{2\alpha}(P\phi_2)}\label{a141}\\
&\le & d_n2^{2\alpha-1}p^{\alpha}(1-p)^{\alpha} \sqrt{P_{\alpha}(\phi_1)P_{\alpha}(\phi_2)}\,.
\ea
In Eq.~\eqref{a14} we used that for any complex number $x$ it holds that $\mathrm{Re}(x)^{2\alpha}+\mathrm{Im}(x)^{2\alpha}\le |x|^{2\alpha}$. For the above chain of inequalities to hold, we need to show Eq.~\eqref{a141}, that is
\be
\sum_{P}\tr^{\alpha}(P\phi_1P\phi_2)\le \sqrt{\sum_{P}\tr^{2\alpha}(P\phi_1)}\sqrt{\sum_P\tr^{2\alpha}(P\phi_2)}\,.\label{toprove100}
\ee
To show Eq.~\eqref{toprove100}, we can first expand each $(n-1)$ qubit state $\phi_i$ in the Pauli basis as $\phi_i=\frac{1}{d_{n-1}}\sum_P\tr(P\phi_i)P$ for $i=1,2$, and write
\be
\sum_{P}\tr^{\alpha}(P\phi_1P\phi_2)&=&\frac{1}{d^{\alpha}_{n-1}}\sum_P\sum_{P_1,\ldots, P_{\alpha}}\prod_{i=1}^{\alpha}[\tr(P_i\phi_1)\tr(P_i\phi_2)]\prod_{i=1}^{\alpha}\Omega(P,P_i)\,,\label{a7}
\ee
where, in Eq.~\eqref{a7}, we defined the following function over Pauli operators:
\be
\Omega(P,Q)\coloneqq \frac{1}{d_{n-1}}\tr(PQP^{\dag}Q^{\dag})=(-1)^{\delta_{QP\neq PQ}}\,,
\ee
which captures the commutation relation between $P$ and $Q$. Notice that the following identity holds
\be
\Omega(P,Q_{1})\Omega(P,Q_{2})=\Omega(P,Q_{1}Q_{2})=d_{n-1}^{-1}\tr(PQ_1Q_2PQ_2Q_1),\quad \forall P,Q_1,Q_2\in\mathbb{P}_{n-1}\,.
\ee
Starting from Eq.~\eqref{a7}, we can use multiple times the above identity and write:
\be
\sum_P\prod_{i=1}^{\alpha}\Omega(P,P_i)=\sum_P\Omega(P,P_{1}\cdots P_{\alpha})=\frac{1}{d_{n-1}}\sum_P\tr(PP_{1}\cdots P_{\alpha}P(P_{1}\cdots P_{\alpha})^{\dag})=d_{n-1}^2\delta_{P_1P_2\dots P_\alpha\propto\mathbb{1}}\,.\label{identity1}
\ee
The last equality follows from Haar integration and the fact that the Pauli group forms a $1$-design. Plugging the identity in Eq.~\eqref{identity1} back to Eq.~\eqref{a7}, we arrive to
\ba
\sum_{P}\tr^{\alpha}(P\phi_1P\phi_2)&=&\frac{1}{d_{n-1}^{\alpha-2}}\sum_{P_1,\ldots,P_\alpha}\delta_{P_1P_2\dots P_\alpha\propto\mathbb{1}}\tr(P_1\phi_1)\tr(P_1\phi_2)\cdots \tr(P_\alpha\phi_1)\tr(P_\alpha\phi_2)\label{a8}\\
&\le& \frac{1}{d_{n-1}^{\alpha-2}}\sqrt{\sum_{P_1,\ldots, P_{\alpha}}\delta\prod_{i=1}^{\alpha}\tr^{2}(P_i\phi_1) }\sqrt{\sum_{P_1,\ldots, P_{\alpha}}\delta\prod_{i=1}^{\alpha}\tr^{2}(P_i\phi_2) }\label{a9}\\
&= &  \sqrt{\frac{1}{d_{n-1}^{\alpha}}\sum_{P,P_1,\ldots, P_{\alpha}}\prod_{i=1}^{\alpha}\tr^{2}(P_i\phi_1)  \prod_{i=1}^{\alpha}\Omega(P,P_i)}\sqrt{\frac{1}{d_{n-1}^{\alpha}}\sum_{P,P_1,\ldots, P_{\alpha}}\prod_{i=1}^{\alpha}\tr^{2}(P_i\phi_2)\prod_{i=1}^{\alpha}\Omega(P,P_i) }\label{a10}\\
&=&\sqrt{\sum_P\tr^{\alpha}(P\phi_1P\phi_1)}\sqrt{\sum_P\tr^{\alpha}(P\phi_2P\phi_2)}\\
&=&\sqrt{\sum_P\tr^{2\alpha}(P\phi_1)}\sqrt{\sum_P\tr^{2\alpha}(P\phi_2)}\,.\label{b20}
\ea
In Eq.~\eqref{a9}, we used the Cauchy–Schwarz inequality for the $d_{n-1}^{2\alpha}$-dimensional vectors with components (for $i=1,2$) 
\be
\delta_{P_1P_2\cdots P_{\alpha}\propto \mathbb{1} }\tr(P_1\phi_i)\times\tr(P_2\phi_i)\times \ldots\times \tr(P_{\alpha}\phi_i), \quad  P_{1},\ldots,P_{\alpha}\in\mathbb{P}\,.
\ee
Moreover, we used a short notation for $\delta\equiv\delta_{P_1P_2\cdots P_{\alpha}\propto \mathbb{1} }$ and we used $\delta^2=\delta$. In Eq.~\eqref{a10} we applied the identity in Eq.~\eqref{identity1} again. Finally, in Eq.~\eqref{b20} we used that $\tr(P\phi_i P\phi_i)=\tr^2(P\phi_i)$ for $i=1,2$.  This proves Eq.~\eqref{toprove100}. Plugging everything back to Eq.~\eqref{A2} and~\eqref{A3}, we have
\be
P_{\alpha}(\psi)&\le \sum_{i=0}^{\alpha}\binom{2\alpha}{2i}p^{2i}(1-p)^{2(\alpha-i)}P_{\alpha}^{\frac{i}{\alpha}}(\phi_1)P_{\alpha}^{\frac{\alpha-i}{\alpha}}(\phi_2)+2^{2\alpha-1}p^{\alpha}(1-p)^{\alpha}\sqrt{P_{\alpha}(\phi_1)P_{\alpha}(\phi_2)}\,,
\ee
which concludes the proof of Eq.~\eqref{app:firstexpressionlemma2}. To obtain Eq.~\eqref{app:secondexpressionlemma2} it is sufficient to perform the sum. 
\end{proof}
\end{lemma}

\begin{corollary}\label{cor:stabentropymin}
Consider $\ket{\psi}=\sqrt{p}\ket{0}\otimes\ket{\phi_1}+\sqrt{1-p}\ket{1}\otimes \ket{\phi_2}$ and $\ket{\phi_1},\ket{\phi_2}\in \mathbb{C}^{2\otimes (n-1)}$. For any integer $\alpha\ge 2$, it follows
\be
M_{\alpha}(\psi)\ge \min\{M_{\alpha}(\phi_1),M_{\alpha}(\phi_2)\}\,.
\ee
\begin{proof}
Starting from Eq.~\eqref{app:firstexpressionlemma2} in \cref{app:lemma2}, we can bound
\be
P_{k}(\psi)\le\left( \sum_{i=0}^{\alpha}\binom{2\alpha}{2i}p^{2i}(1-p)^{2(\alpha-i)}+2^{2\alpha-1}p^{\alpha}(1-p)^{\alpha}\right)\max\{P_{\alpha}(\phi_1),P_{\alpha}(\phi_2)\}\,.
\ee
The coefficient reads
\be
\sum_{i=0}^{\alpha}\binom{2\alpha}{2i}p^{2i}(1-p)^{2(\alpha-i)}+2^{2\alpha-1}p^{\alpha}(1-p)^{\alpha}=\frac{1}{2} \left(\left(2 p-1\right)^{2 \alpha}+p^{ \alpha} \left(4-4 p\right)^\alpha+1\right)\le 1\,.
\ee
Therefore, one has $P_{k}(\psi)\le \max\{P_{\alpha}(\phi_1),P_{\alpha}(\phi_2)\}$ which implies the statement (see Definition 4).
\end{proof}
\end{corollary}

\begin{corollary}\label{cor:strongpurity}
Consider $\ket{\psi}=\sqrt{p}\ket{0}\otimes\ket{\phi_1}+\sqrt{1-p}\ket{1}\otimes \ket{\phi_2}$ and $\ket{\phi_1},\ket{\phi_2}\in \mathbb{C}^{2\otimes (n-1)}$. For any integer $\alpha\ge 2$, it holds that
\be
P_{\alpha}(\psi)\le pP_{\alpha}(\phi_1)+(1-p)P_{\alpha}(\phi_2)\,.
\ee
\begin{proof}
Let us start from Eq.~\eqref{app:firstexpressionlemma2} in \cref{app:lemma2}, which we display below for convenience.
\be
P_{\alpha}(\psi)\le \sum_{i=0}^{\alpha}\binom{2\alpha}{2i}p^{2i}(1-p)^{2(\alpha-i)}P_{\alpha}^{\frac{i}{\alpha}}(\phi_1)P_{\alpha}^{\frac{\alpha-i}{\alpha}}(\phi_2)+2^{2\alpha-1}p^{\alpha}(1-p)^{\alpha}\sqrt{P_{\alpha}(\phi_1)P_{\alpha}(\phi_2)}\,.\label{23}
\ee
First of all, note that $P_{\alpha}^{\frac{i}{\alpha}}(\phi_1)P_{\alpha}^{\frac{\alpha-i}{\alpha}}(\phi_2)\le \frac{i}{\alpha}P_{\alpha}(\phi_1)+\frac{\alpha-i}{\alpha}P_{\alpha}(\phi_2)$ for the geometric-arithmetic mean inequality. We can perform the sum over the index $i$ easily. Since the expression is symmetric in $(p,P_{\alpha}(\phi_1))\leftrightarrow(1-p,P_{\alpha}(\phi_2))$, without loss of generality, let us pose $p\ge1/2$. We thus get:
\be
P_{\alpha}(\psi)\le \frac{1}{2}p[(2p-1)^{2\alpha-1}+1]P_{\alpha}(\phi_1)-\frac{1}{2}(1-p)[(2p-1)^{2\alpha-1}-1]P_{\alpha}(\phi_2)+2^{2\alpha-1}p^{\alpha}(1-p)^{\alpha}\sqrt{P_{\alpha}(\phi_1)P_{\alpha}(\phi_2)}\,.\label{app:eq1}
\ee
Then, to show the statement, it is sufficient to show that the right hand side of Eq.~\eqref{app:eq1} is upper bounded by $pP_{\alpha}(\phi_1)+(1-p)P_{\alpha}(\phi_2)$. Taking the difference between the two terms, we need to show
\be
F(x)=-f(p)x^2+g(p)x-h(p)\le 0,\quad x\ge 0\,,
\ee
for every $1/2\le p\le1$, where $f(p)=\frac{p}{2}(1-(2p-1)^{2\alpha-1})$, $h(p)=\frac{1-p}{2}((2p-1)^{2\alpha-1}+1)$ and $g(p)=2^{2\alpha-1}p^{\alpha}(1-p)^{\alpha}$. Note that we defined $x\coloneqq \sqrt{\frac{P_{\alpha}(\phi_1)}{P_{\alpha}(\phi_2)}}$\,.

For $x=0$ and $x\rightarrow\infty$ it readily holds that $F(x)\le 0$ since $f(p),h(p)\ge 0$. Maximizing over $x$, we get $F'(x)=g(p)-2f(p)x$, and substituting $x(p)=\frac{g(p)}{2f(p)}$ into $F(x)$, we get
\be
F_{\alpha}(x(p))=\frac{(1-p)}{2}\left(-[1+(2p-1)^{2\alpha-1}]+\frac{[4p(1-p)]^{2\alpha-1}}{1-(2p-1)^{2\alpha-1}}\right)\,.
\ee
We are left to show that $F_{\alpha}(x(p))\le 0$ for all $1/2\le p\le 1$. Except for $p=1$ for which $F_{\alpha}(x(1))=0$ and the result readily follows, this is equivalent to show that $\tilde{F}_{\alpha}(p)\le 0$, where
\be
\tilde{F}_{\alpha}(p)=-[1-(2p-1)^{2(2\alpha-1)}]+[4p(1-p)]^{2\alpha-1}\,.
\ee
Noting that $\tilde{F}_{\alpha}$ is a monotonically decreasing function in $\alpha$ (being the sum of two monotonically decreasing functions. Indeed note that $4p(1-p)\le 1$), it is sufficient to show the statement for $\alpha=2$. We have
\be
\tilde{F}_2(p)=[4p(1-p)]^{3}-[1-(2p-1)^{6}]=-12p(1-p)(2p-1)^2\le 0\,,
\ee
which concludes the proof.
\end{proof}
\end{corollary}

\section{Proof of main results: Theorem 1, Theorem 2 and Theorem 3}\label{appB}

{\bf Proof of Theorem 1.} Let us recall the definition of pure-state magic monotone. It must hold that for every pair $(\ket{\psi},\mathcal{E})$ where $\mathcal{E}\in\mathcal{S}$ obeying $\mathcal{E}(\ketbra{\psi})=\ketbra{\phi}$, then $M_{\alpha}(\psi)\ge M_{\alpha}(\phi)$. To show pure-state monotonicity of stabilizer entropies with $\alpha\ge2$, we make use of \cref{cor:stabentropymin}. First of all, any stabilizer protocol can be decomposed as a combination of elementary operations, i.e., (i) Clifford unitaries; (ii) partial trace; (iii) measurements in the computational basis; (iv) composition with ancillary qubits in $\ket{0}$; the above operations conditioned on (v) measurement outcomes and on (vi) classical randomness. 

It is crucial to note that a deterministic pure-state stabilizer protocol only necessitates the initial and final states to be pure. However, given that a general stabilizer protocol consists of, let us say, $N$ elementary operations, denoted as $\mathcal{E}=\mathcal{E}_{N}\circ \mathcal{E}_{N-1}\circ\cdots\circ\mathcal{E}_1$, there is no obligation for each step to maintain a pure state, nor must it be deterministic. Therefore, given a collection of (possibly mixed) states $\{(p_i,\rho_i)\}$, we define
\be
M_{\alpha}^{\min}(\{(p_i,\rho_i\})\coloneqq \inf_{\substack{(q_{ij},\phi_{ij})}}\Big\{\min_{ij}M_{\alpha}(\phi_{ij})\,:\, p_i\rho_i =\sum_jq_{ij} \phi_{ij} \,,\, \forall i\Big\} \,,
\ee
which is a possible extension of the stabilizer entropy $M_{\alpha}$ to a collection of states. To clarify the notation: given all the states $\rho_i$; for each convex decomposition $\rho_i=\sum_jq_{ij}\phi_{ij}$, we choose the minimum  $\min_{ij}M_{\alpha}(\phi_{ij})$ and then optimize over all the possible convex decompositions $\inf_{(q_{ij},\phi_{ij})}$. Therefore, after the optimization, we can describe any state of the system by a collection of pure states $\{(p_iq_{ij}, \phi_{ij})\}$, which describes the optimal decomposition of $\{(p_i,\rho_i)\}$ with respect to $M_\alpha$ and, after the optimization, we have $M_{\alpha}^{\min}(\{(p_i,\rho_i)\})=\min_{ij}\{M_{\alpha}(\phi_{ij})\}$.

The strategy of the proof is as follows: we first demonstrate that $M_{\alpha}^{\min}$ is a magic monotone according to Definition 1 in full generality. Then, by restricting deterministic stabilizer protocols, we establish that $M_{\alpha}$ is a pure-state magic monotone, given that (crucially) $M_{\alpha}(\phi)=M_{\alpha}^{\min}(\phi)$ for every pure state $\phi$.

With this tool at hand, let us first show that, starting from a pure state $\phi$, for every elementary stabilizer operation $\mathcal{E}(\cdot)$ it holds that 
\be
M_{\alpha}(\phi)\ge M_{\alpha}^{\min}(\mathcal{E}(\phi))\,.\label{b11}
\ee
\begin{itemize}
    \item {\em Clifford unitaries and appending stabilizer ancillas.} Eq.~\eqref{b11} holds trivially because $M_{\alpha}$ is invariant under unitary Clifford transformations and it is additive, see \cref{app:stabentropy}.

    \item {\em Measurement in the computational basis.} This follows directly from \cref{cor:stabentropymin}. Since we map the state $\ket{\psi}=\sqrt{p}\ket{0}\otimes\ket{\phi_1}+\sqrt{1-p}\ket{1}\otimes\ket{\phi_2}$ to the collection of pure states $\{(p,\phi_1),(1-p,\phi_2)\}$, for which $M_{\alpha}(\psi)\ge \min\{M_{\alpha}(\phi_1),M_{\alpha}(\phi_2)\}\ge M^{\min}_{\alpha}(\mathcal{E}(\psi))$. If the post measurement state is maintained after measurement, the same follows because $M_{\alpha}(\phi_{i+1})=M_{\alpha}(\ketbra{i}\otimes\phi_{i+1})$ for $i=0,1$.

    \item {\em Partial trace and dephasing.} Both dephasing and partial trace can be simulated by a forgetful measurement.
    We can therefore select $\{(p,\phi_1),(1-p,\phi_2)\}$ for partial trace or $\{(p,\ketbra{0}\otimes \phi_1),(1-p,\ketbra{1}\otimes \phi_2)\}$ as our convex mixtures to upper-bound the infimum. This allows us to always maintain a pure state collection $\{(p_i,\phi_i)\}$ for every operation.

\item {\em Conditional operations based on measurement outcomes or classical randomness.} In general, any elementary conditional operation acts as
\begin{align}
    \phi\mapsto\{(p_i,\phi_i)\}\mapsto \{(p_i,\mathcal{E}_i(\phi_i))\}\,,
\end{align}
where the first map comes from the result of a computational basis measurement on $\phi$ or a classical randomness splitting, i.e. $\phi\mapsto \{(p,\phi),(1-p,\phi)\}$. Indeed note that this is the most general form of classical randomness: it includes both conditional preparation of stabilizer states and conditional operations. We have already seen that $M_{\alpha}(\phi)\ge M_{\alpha}^{\min}(\{(p_i,\phi_i)\})$ for a measurement. While, trivially, $M_{\alpha}(\phi)\ge M_{\alpha}(\phi)$ for classical randomness. Moreover, for every $i$, $M_{\alpha}(\phi_i)\ge M_{\alpha}^{\min}(\mathcal{E}_i(\phi_i))$. Hence:
\begin{align}
    M_{\alpha}(\phi)\ge M_{\alpha}^{\min}(\{(p_i,\phi_i)\})=\min_i M_{\alpha}(\phi_i)\geq \min_i M^{\min}_{\alpha}(\mathcal{E}_i(\phi_i))\ge M_{\alpha}^{(\min)}(\{p_i,\mathcal{E}_i(\phi_i)\})\,.
\end{align}
The second equality follows by definition, while the last inequality follows because we have chosen a particular decomposition, not necessarily the infimum. The result just follows for conditional operations.

\end{itemize}
Now, we generalize the result to collections of mixed states and show that:
\begin{align}
    M^{\min}_{\alpha}(\{(p_i,\rho_i)\})\ge M^{\min}_{\alpha}(\mathcal{E}(\{(p_i,\rho_i)\}))\,.\label{toshow1}
\end{align}
Let $\{(q_{ij},\phi_{ij})\}$ be the optimal decomposition of $\{(p_i,\rho_i)\}$ into an collection of pure states. Consider an elementary stabilizer operation $\mathcal{E}$. We have already shown that, for every state $\phi_{ij}$ of such a collection, it holds that
\be
M_{\alpha}(\phi_{ij})\ge M^{\min}_{\alpha}(\mathcal{E}(\phi_{ij}))\,.
\ee
One can also verify that $\{(q_{ij}, \mathcal{E}(\phi_{ij}))\}$ is a potentially mixed state collection of $\mathcal{E}(\{(p_i,\rho_i)\})$. It therefore upper bounds the optimal value. We therefore have:
\begin{align}
   M^{\min}_\alpha(\mathcal{E}(\{(p_i,\rho_i)\}))\leq\min_{ij} M^{\min}_\alpha(\mathcal{E}(\phi_{ij}))\leq \min_{ij} M_\alpha(\phi_{ij})=M_\alpha^{\min}(\{(p_i,\rho_i)\})\,.
\end{align}
The first inequality arises since we selected a particular decomposition, i.e., $\{(q_{ij}, \mathcal{E}(\phi_{ij}))\}$, which is not necessarily the infimum; the last equality follows from $\{(q_{ij},\phi_{ij})\}$ being a optimal decomposition of $\{(p_i,\rho_i)\}$. Since Eq.~\eqref{toshow1} holds for all elementary stabilizer operations, it also holds for all composite stabilizer operations. To conclude the proof, consider a deterministic stabilizer protocol $\mathcal{E}$ that, starting from $\ketbra{\psi}$ returns a pure state $\ketbra{\phi}$. We therefore have $M_{\alpha}(\psi)\ge M_{\alpha}^{\min}(\phi)=M_{\alpha}(\phi)$, which concludes the proof. Moreover, if the stabilizer protocol $\mathcal{E}$ is non-deterministic and yields a collection of pure-states $\mathcal{E}(\psi)=\{(p_i,\phi_i)\}$, we have $M_{\alpha}(\psi)\ge\min_iM_{\alpha}(\phi_i)$.

\medskip

{\bf Proof of Theorem 3.} Let us first recall the definition of extended stabilizer purity $\widehat{P}_{\alpha}$ on a collection of (possibly mixed) states $\{(p_i,\rho_i)\}$:
\be
\widehat{P}_{\alpha}(\{(p_i,\rho_i\})\coloneqq \sup_{\substack{(q_{ij},\phi_{ij})}}\Big\{\sum_{ij}q_{ij}P_{\alpha}(\phi_{ij})\,:\, p_i\rho_i =\sum_jq_{ij} \phi_{ij}\,,\forall i\Big\} \,.\label{b1}
\ee
That is, given the collection of states, we first compute the average stabilizer purity $P_{\alpha}$ defined on pure states in Definition 4 with respect to a convex decomposition of $p_i\rho_i$ and then take the superior over all the possible convex decompositions. 
We will use the same strategy used for the proof of Theorem 1: we first show that for every pure state $\phi$ and elementary stabilizer operation $\mathcal{E}$, the following holds:
\be
P_{\alpha}(\phi)\le \widehat{P}_{\alpha}(\mathcal{E}(\phi))\,.\label{c9}
\ee
\begin{itemize}
    \item {\em Clifford unitaries and appending stabilizer ancillas.} Eq.~\eqref{c9} follows because $P_{\alpha}$ is invariant under Clifford unitaries and it is multiplicative, see \cref{app:stabentropy}.
    \item {\em Measurement in the computational basis.} This follows directly from \cref{cor:strongpurity}. Since we map the state $\ket{\phi}=\sqrt{p}\ket{0}\otimes\ket{\phi_1}+\sqrt{1-p}\ket{1}\otimes\ket{\phi_2}$ to the collection of pure states $\{(p,\phi_1),(1-p,\phi_2)\}$, for which $P_{\alpha}(\phi)\le pP_{\alpha}(\phi_1)+(1-p) P_{\alpha}(\phi_2)$ holds. If the post measurement state is maintained after measurement, the same follows since $P_{\alpha}(\phi_{i+1})=P_{\alpha}(\ketbra{i}\otimes\phi_{i+1})$ for $i=0,1$.
    \item {\em Partial trace and dephasing.} 
    Both dephasing and partial trace can be simulated by a forgetful measuremnt.
    We can therefore select $\{(p,\phi_1),(1-p,\phi_2)\}$ for partial trace or $\{(p,\ketbra{0}\otimes \phi_1),(1-p,\ketbra{1}\otimes \phi_2)\}$ as our convex mixture to lower-bound the superior. This allows us to always maintain a pure state collection $\{(p_i,\phi_i)\}$ for every operation.

\item {\em Conditional operations based on measurement outcomes or classical randomness.} In general, any elementary conditional operation acts as
\begin{align}
    \phi\mapsto\{(p_i,\phi_i)\}\mapsto \{(p_i,\mathcal{E}_i(\phi_i))\}\,,
\end{align}
where the first map comes from the result of a computational basis measurement on $\phi$ or a classical randomness splitting, i.e. $\phi\mapsto \{(p,\phi),(1-p,\phi)\}$. Indeed note that this is the most general form of classical randomness: it includes both conditional preparation of stabilizer states and conditional operations. We have already seen that $P_{\alpha}(\phi)\le \widehat{P}_{\alpha}\{(p_i,\phi_i)\})=\sum_{i}p_iP_{\alpha}(\phi_i)$ for a measurement. While, somewhat trivially, it holds that $P_{\alpha}(\phi)\le pP_{\alpha}(\phi)+(1-p)P_{\alpha}(\phi)$ for classical randomness. Moreover we have also shown that, for every $i$, $P_{\alpha}(\phi_i)\le \widehat{P}_{\alpha}(\mathcal{E}_i(\phi_i))$. Hence:
\begin{align}
    P_{\alpha}(\phi)\le \widehat{P}_{\alpha}(\{(p_i,\phi_i)\})=\sum_ip_i P_{\alpha}(\phi_i)\leq \sum_i p_i \widehat{P}_{\alpha}(\mathcal{E}_i(\phi_i))\le \widehat{P}_{\alpha}(\{(p_i,\mathcal{E}_i(\phi_i))\})\,.
\end{align}
The first equality follows by definition: note that for a collection of pure states, no optimization is needed. The last inequality follows because we have chosen a particular decomposition, not necessarily the superior. The result just follows for conditional operations. 
\end{itemize}
Now, we generalize the above result to collections of mixed states and show that for every collection $\{(p_i,\rho_i)\}$ and elementary stabilizer operation $\mathcal{E}$, it holds that:
\begin{align}
    \widehat{P}_{\alpha}(\{(p_i,\rho_i)\})\le \widehat{P}_{\alpha}(\mathcal{E}(\{(p_i,\rho_i)\}))\,.\label{toshow2}
\end{align}
Let $\{(q_{ij},\phi_{ij})\}$ be the optimal decomposition of $\{(p_i,\rho_i)\}$ into an collection of pure states. Consider an elementary stabilizer protocol $\mathcal{E}$. We have already shown that, for every pure state $\phi_{ij}$ in such a collection, it holds that
\be
\widehat{P}_{\alpha}(\mathcal{E}(\phi_{ij}))\ge P_{\alpha}(\phi_{ij})\,.
\ee
One can also verify that $\{(q_{ij}, \mathcal{E}(\phi_{ij}))\}$ is a potentially mixed state collection of $\mathcal{E}(\{(p_i,\rho_i)\})$. Therefore, we can write the following chain of inequalities:
\begin{align}
   \widehat{P}_{\alpha}(\mathcal{E}(\{(p_i,\rho_i)\}))\ge\sum_{ij}  q_{ij}\widehat{P}_{\alpha}(\mathcal{E}(\phi_{ij}))\geq \sum_{ij}  q_{ij} P_{\alpha}(\phi_{ij})=\widehat{P}_{\alpha}(\{(p_i,\rho_i)\})\,.
\end{align}
where the first inequality arises since we selected a particular decomposition, i.e., $\{(q_{ij}, \mathcal{E}(\phi_{ij}))\}$, which is not necessarily the superior. The last equality follows because $\{(q_{ij}, \mathcal{E}(\phi_{ij}))\}$ is an optimal decomposition. Since this holds for all elementary stabilizer operations, Eq.~\eqref{toshow2} holds every stabilizer protocol. 

Lastly, the monotonicity of the extended stabilizer entropy and linear stabilizer entropy readily follows from Eq.~\eqref{toshow2}. What's more, the above theorem also shows that linear stabilizer entropies are strong monotones not only restricted to pure-states. 

\medskip
{\bf Proof of Theorem 2.} The proof of Theorem 2 readily follows from Theorem 3. To see this, let us recall the definition of pure-state strong monotone. Starting from a pure state $\phi$, for any non-deterministic stabilizer protocol $\mathcal{E}$ ending in a collection $ \mathcal{E}\,:\, \phi\mapsto\{(p_i,\phi_i)\}$ of pure states $\phi_i$, it must hold that $\mathcal{M}(\phi)\ge \sum_ip_i\mathcal{M}(\phi_i)$. Given the definition of extended linear stabilizer entropy, we have shown that
\be
M_{\alpha}^{\mathrm{lin}}(\phi)=\widehat{M}_{\alpha}^{\mathrm{lin}}(\phi)\ge \widehat{M}_{\alpha}^{\mathrm{lin}}(\{(p_i,\phi_i)\})=\sum_ip_iM_{\alpha}^{\mathrm{lin}}(\phi_i)\,.
\ee
The first equality follows because the two definitions adhere for pure states. The inequality follows because $\widehat{M}_{\alpha}^{\mathrm{lin}}$ is a magic monotone thanks to Theorem 3, and the last equality follows by definition (see Definition 5). Note indeed that there is no optimization over convex decompositions being $\phi_i$ pure states. This  concludes the proof. 
\section{Improved resource conversion bounds}
In this section, we will exploit the established monotonicity of stabilizer entropies to provide improved and tight bounds for resource conversion, as explained in the main text. In particular, starting from the consideration of~\cite{beverland_lower_2020}, we improve upon the provided bounds. Being not the main focus of this paper, we limit the analysis to improve resource conversion between the $T$-state $\ket{T}$ and the following $n$-qubit states
\ba
\ket{C^{n-1}Z}&=&\frac{1}{\sqrt{2^n}}\sum_{b\in\{0,1\}^n}(-1)^{b_1\cdots b_n}\ket{b}\\
\ket{C^{n-1}S}&=&\frac{1}{\sqrt{2^n}}\sum_{b\in\{0,1\}^n}(i)^{b_1\cdot b_2\cdots b_n}\ket{b}
\ea
In order to derive conversion bounds, let us compute the $\alpha$-stabilizer purity for these states. While the $\alpha$-stabilizer purity for $\ket{T}$ has been computed in~\cite{leone_stabilizer_2022} and reads
\be
P_{\alpha}(\ket{T})=\frac{2^{\alpha}+2}{2^{\alpha+1}}
\ee
in what follows, we explicitly compute the $\alpha$-stabilizer purity for $\ket{C^{n-1}Z}$ and $\ket{C^{n-1}S}$ respectively.
\begin{proposition}
    The $\alpha$-stabilizer purity for $\ket{C^{n-1}Z}$ and $\ket{C^{n-1}S}$ reads
    \ba
    P_{\alpha}(\ket{C^{n-1}Z})&=&P_{\alpha}(\ket{C^{n-1}Z})=\frac{1}{d_n}\left(1+\left(1-\frac{4}{d_n}\right)^{2\alpha}(d_n-1)+\frac{2^{\alpha-1}}{d^{2\alpha}_n}(d^{2}_n-3d_n+2)\right)\label{stabpurityCCZ}\\
    P_\alpha(\ket{C^{n-1}S})&=&\frac{1}{d_n}\left(1+\left(1-\frac{2}{d_n}\right)^{2\alpha}(d_n-1)+\frac{2^{2\alpha}}{d_{n}^{2\alpha}}(d_n-1)^2\right)\label{stabpurityCCS}
    \ea
\begin{proof}
Eq.~\eqref{stabpurityCCZ} is a direct corollary of Proposition 4.2 in Ref.~\cite{beverland_lower_2020}, in which the authors compute all the Pauli expectation values explicitly. Following a similar strategy of Ref.~\cite{beverland_lower_2020}, we now compute all the Pauli expectation values for $\ket{C^{n-1}S}$. 
Since we can express every Pauli, up to phase, as $P\propto X^xZ^z$ for $x,z\in\{0,1\}^n$, we can compute the expectation value of X and Zs products, i.e. $\langle C^{n-1}S|X^xZ^z|C^{n-1}S\rangle$,  and then consider the absolute value. By direct computation we have
\ba
2^n\langle C^{n-1}S|X^xZ^z|C^{n-1}S\rangle&=&\sum_{b,b^{\prime}\in\{0,1\}^n}(i)^{b_1\cdot b_2\cdots b_n}(-i)^{b_{1}^{\prime}\cdot b_{2}^{\prime}\cdots b_{n}^{\prime}}\langle b^{\prime}|X^xZ^z|b\rangle \\
&=&\sum_{b,b^{\prime}\in\{0,1\}^n}(i)^{b_1\cdot b_2\cdots b_n}(-i)^{b_{1}^{\prime}\cdot b_{2}^{\prime}\cdots b_{n}^{\prime}}(-1)^{b\cdot z}\langle b^{\prime}|b+x\rangle \\
&=&\sum_{b\in\{0,1\}^n}(i)^{b_1\cdot b_2\cdots b_n}(-i)^{(b_{1}+x_1)\cdot (b_{2}+x_n)}(-1)^{b\cdot z}
\ea
For $x=0^n$, we have that the sum reduces to $\sum_{b\in\{0,1\}^n}(-1)^{b\cdot z}=2^n\delta_{z,0^n}$. For every $x\neq 0^n$, thanks to the 'and' structure, we have a result different from $\sum_{b\in\{0,1\}^n}(-1)^{b\cdot z}$ only for $b=1^n$ or $b=x+1^n$, hence we can write (for $x\neq 0^n$)
\ba
\sum_{b\in\{0,1\}^n}(i)^{b_1\cdot b_2\cdots b_n}(-i)^{(b_{1}+x_1)\cdot (b_{2}+x_n)}(-1)^{b\cdot z}&=&i(-1)^{z\cdot 1^n}-i(-1)^{z\cdot (1^n+x)}+\sum_{\substack{b\in\{0,1\}^n\\b\neq 1^n,1^n+x}}(-1)^{b\cdot z}\\
&=&(i-1)(-1)^{z\cdot 1^n}+(-i-1)(-1)^{z\cdot (1^n+x)}+\sum_{b\in\{0,1\}^n}(-1)^{b\cdot z}\\
&=&(i-1)(-1)^{z\cdot 1^n}+(-i-1)(-1)^{z\cdot (1^n+x)}+2^n\delta_{z,0^n}
\ea
For $z=0^n$, then it reduces to $2^n-2$. Now, for any $z\neq 0^n$, it is $(i-1)(-1)^{z\cdot 1^n}+(-i-1)(-1)^{z\cdot (1^n+x)}$. We have just to discriminate 2 cases. When $x\cdot z=0\mod 2$, it returns $\pm 2$ depending on $(-1)^{z\cdot 1^n}$. While, for $x\cdot z=1\mod 2$, it returns $\pm 2i$, depending on $(-1)^{z\cdot 1^n}$. Let us list what we have got by considering the absolute value $|\langle C^{n-1}S|X^xZ^z|C^{n-1}S\rangle|$. 
\be
|\langle C^{n-1}S|X^xZ^z|C^{n-1}S\rangle|=\begin{cases}
1\quad \quad\quad\quad\text{if}\, x=0^n\, \text{and}\, z=0^n\\
0\quad \quad\quad\quad\text{if}\, x=0^n\, \text{and}\, z\neq 0^n\\
1-2^{1-n}\quad \text{if}\, x\neq 0^n\,\text{and}\, z= 0^n\\
2^{1-n}\quad\quad\quad \text{if}\, x\neq 0^n\,\text{and}\, z\neq0^n\\
\end{cases}
\ee
Counting the various cases, we can easily compute the $\alpha$-stabilizer purity as
\be
P_\alpha(\ket{C^{n-1}S})=\frac{1}{d_n}\left(1+\left(1-\frac{2}{d_n}\right)^{2\alpha}(d_n-1)+\frac{2^{2\alpha}}{d_{n}^{2\alpha}}(d_n-1)^2\right)
\ee
which concludes the proof. 
\end{proof}
\end{proposition}
Using the additivity property of stabilizer entropies (see Sec.~\ref{app:stabentropy}), we can now bound the asymptotic rate of conversion for the considered resource states~\cite{beverland_lower_2020}. 

\medskip
{\bf Conversion rate bounds.} While in the main text, we report only a part of our results, in the following, we bound the conversion rate of the following transformation $\ket{C^{n-1}S}\mapsto\ket{\phi}$ and $\ket{C^{n-1}Z}\mapsto\ket{\phi}$ for $\ket{\phi}=\ket{T},\ket{CS},\ket{CCZ}$ improving on the current state of the art bounds. 
\begin{itemize}
    \item $r[\ket{C^{n-1}Z}\mapsto\ket{T}]\le \frac{M_{2}(\ket{C^{n-1}Z})}{M_{2}(\ket{T})}=-\frac{\log \left(16^{-n} \left(-7\ 2^{n+5}-2^{3 n+4}+7\ 4^{n+2}+16^n+128\right)\right)}{\log \left(\frac{4}{3}\right)}$;
    \item $r[\ket{C^{n-1}Z}\mapsto\ket{CS}]\le \frac{M_{2}(\ket{C^{n-1}Z})}{M_{2}(\ket{CS})}=-\frac{\log \left(16^{-n} \left(-7\ 2^{n+5}-2^{3 n+4}+7\ 4^{n+2}+16^n+128\right)\right)}{\log \left(\frac{16}{7}\right)}$;
    \item $r[\ket{C^{n-1}Z}\mapsto\ket{CCZ}]\le \frac{M_{2}(\ket{C^{n-1}Z})}{M_{2}(\ket{CCZ})}=-\frac{\log \left(16^{-n} \left(-7\ 2^{n+5}-2^{3 n+4}+7\ 4^{n+2}+16^n+128\right)\right)}{\log \left(\frac{32}{11}\right)}$;
    \item $r[\ket{C^{n-1}S}\mapsto\ket{T}]\le \frac{M_{2}(\ket{C^{n-1}S})}{M_{2}(\ket{T})}=-\frac{\log \left(2^{-4 n} \left(-5\ 2^{n+3}+2^{2 n+5}-8^{n+1}+16^n+16\right)\right)}{\log \left(\frac{4}{3}\right)}$;
    \item $r[\ket{C^{n-1}S}\mapsto\ket{CS}]\le \frac{M_{2}(\ket{C^{n-1}S})}{M_{2}(\ket{CS})}=-\frac{\log \left(2^{-4 n} \left(-5\ 2^{n+3}+2^{2 n+5}-8^{n+1}+16^n+16\right)\right)}{\log \left(\frac{16}{7}\right)}$;
    \item $r[\ket{C^{n-1}S}\mapsto\ket{CCZ}]\le \frac{M_{2}(\ket{C^{n-1}S})}{M_{2}(\ket{CCZ})}=-\frac{\log \left(2^{-4 n} \left(-5\ 2^{n+3}+2^{2 n+5}-8^{n+1}+16^n+16\right)\right)}{\log \left(\frac{32}{11}\right)}$;
\end{itemize}
In particular, we improve upon the bounds presented in Ref.~\cite{beverland_lower_2020} Table 2, for which we obtain
\begin{itemize}
    \item $r[\ket{C^3Z}\mapsto \ket{CCZ}]\le 0.9$
    \item $r[\ket{C^4Z}\mapsto \ket{CCZ}]\le 0.5$
    \item $r[\ket{C^2S}\mapsto \ket{CCZ}]\le 0.8$ 
    \item $r[\ket{C^3S}\mapsto \ket{CCZ}]\le 0.5 $

\end{itemize}

\end{document}